\documentclass[11pt,letterpaper]{article}
\usepackage[margin=0.5in]{geometry}
\usepackage{amssymb}
\usepackage{amsmath}
\usepackage{pifont}
\usepackage{array}
\usepackage{wrapfig}
\usepackage{graphicx}
\usepackage{url}
\usepackage{cite}
\usepackage{caption}
\usepackage{subcaption}
\usepackage{tablefootnote}

\usepackage[labelfont=bf,figurename=Fig]{caption}

\title{Energy-Efficient Approximate Full Adders Applying Memristive Serial IMPLY Logic For Image Processing\thanks{Preprint Submitted to arXiv}}

\author{Seyed Erfan Fatemieh$^{1,}$\thanks{Corresponding author: Seyed Erfan Fatemieh (erfanfatemieh@eng.ui.ac.ir)}~, and Mohammad Reza Reshadinezhad$^{1,}$\thanks{Corresponding author: Mohammad Reza Reshadinezhad (m.reshadinezhad@eng.ui.ac.ir)}}

\begin{document}
\maketitle

{\noindent$^{1}$ \small \textit{Department of Computer Architecture, Faculty of Computer Engineering, University of Isfahan, Isfahan 8174673441, Iran}}
\section*{Abstract}
Researchers and designers are facing problems with memory and power walls, considering the pervasiveness of Von-Neumann architecture in the design of processors and the problems caused by reducing the dimensions of deep sub-micron transistors. Memristive Approximate Computing (AC) and In-Memory Processing (IMP) can be promising solutions to these problems. We have tried to solve  the power and memory wall problems by presenting the implementation algorithm of four memristive approximate full adders applying the Material Implication (IMPLY) method. The proposed circuits reduce the number of computational steps by up to 40\% compared to the state-of-the-art. The energy consumption of the proposed circuits improves over the previous exact ones by 49\%-75\% and over the approximate full adders by up to 41\%. Multiple error evaluation criteria evaluate the computational accuracy of the proposed approximate full adders in three scenarios in the 8-bit approximate adder structure. The proposed approximate full adders are evaluated in three image processing applications in three scenarios. The results of application-level simulation indicate that the four proposed circuits can be applied in all three scenarios, considering the acceptable image quality metrics of the output images.

\subsection*{Keywords}
Approximate Computing, Approximate Full Adder, Memristor, IMPLY logic, Image Processing, In-Memory Processing.

\section{Introduction} \label{sec1}
The power wall is a considerable adversity in the computer architects' path \cite{ref1}. Energy consumption in portable electronic devices becomes a significant challenge with the invalidation of Dennard's scaling, deviation from Moore's law, and the occurrence of problems with the reduction of transistor dimensions, such as increased leakage currents and short channel effects
\cite{ref2,ref3,ref4,ref5,ref6,ref7,nref2, reviewer23}. The growing difference in the speed of memories (the rate of data that can be supplied to the processor from main memory) and processors known as Von-Neumann’s memory wall bottleneck is another challenge in the design of digital systems \cite{ref1,ref8,ref9,ref10}.

Several solutions to overcome the power wall problem have been proposed to date. One of the solutions welcomed by researchers and technology companies like IBM in recent years is Approximate Computing (AC) \cite{ref2,ref3,ref4,ref11,ref12,ref13}. In AC, the accuracy of computations is reduced to an acceptable level. Designers can reduce circuit complexity as a result of this accuracy reduction \cite{reviewer21}. 
Error-resistant applications such as image processing and 
machine learning 
are among the applications in which AC can be applied \cite{ref2,ref4,ref13,ref14,ref15,ref16,reviewer22}. For example, in image processing, reasons such as a small error in the output pixels due to data noise or approximately performing computations affect the quality of the images \cite{ref2,ref14,ref16}. However, humans' limited visual perception accepts the results \cite{ref2,ref3,ref4,ref16,ref17}.

The idea of applying processing units next to the main memory 
has been applied by researchers for decades to overcome the memory wall. 
With the advancement of technology and the introduction of memristors, it is possible to apply In-Memory Processing (IMP) more than ever before. A memristor is an electrical element that maintains its previous resistance value when voltage and current are not applied to it and is contemplated as a memory cell \cite{ref18,ref19,ref20}. As a memory element and in stateful logic, the High Resistance State (HRS) of a memristor is considered as logic `0', and its Low Resistance State (LRS) is considered as logic `1' \cite{ref3,ref9,ref13,ref21}. Different memory technologies have been contemplated for IMP. However, Resistive RAM (ReRAM) is a popular option among emerging memory technologies for IMP due to its small size, suitable writing speed, low energy consumption, and acceptable endurance \cite{ref10,ref18,ref22,ref23}. ReRAMs fit well in crossbar array structure and harmonize entirely with current manufacturing technology \cite{ref3,ref9,ref18}. There are several methods for designing computing structures that apply memristors. Among the most important of these methods, Memristor Aided loGIC (MAGIC), fast and energy-efficient logic in memory (FELIX), and Material Implication (IMPLY) are attractive \cite{ref10,ref18,ref20,ref23,ref24,ref25}.

Arithmetic circuits are one of the most critical units of each processor. Adders are the central pillar of computing circuits. For example, many instructions in Digital Signal Processing (DSP) processors require adder and multiplier units \cite{ref3,ref26}. So far, many approximate adders have been introduced in different technologies for error-resilient applications that fit today's typical architecture \cite{ref3,ref4,ref12,ref13,ref14,ref15,ref17,ref27,ref28,ref29}. In \cite{ref3,ref13, nref1,nref3}, approximate memristor-based full adders are proposed by applying the IMPLY logic for IMP. In this article, AC and IMP are applied alongside by memristors and IMPLY method, and four approximate full adders are proposed for error-resilient applications. The contributions made in this article are as follows:
\begin{enumerate}
	\item Proposing a compact approximate full adder with Exact $Carry~out$ ($C_{out}$) and Inexact $Sum$ (ECIS) to prevent the propagation of the wrong carry digit from the Least Significant Bits (LSBs) to the Most Significant Bits (MSBs) and improve energy consumption and the number of computational steps compared to the exact full adders,
	\item Presentation of three compact approximate full adders with Inexact $C_{out}$ and Inexact $Sum$ (ICIS1-3) to reduce the number of computational steps and energy consumption compared to the previous designs,
	\item Evaluating the accuracy of computations by multiple error evaluation criteria,
	\item Evaluation of the proposed approximate full adders in three image processing applications to ensure the accuracy of computations in error-resilient applications in three different scenarios,
	\item Presenting two Figures of Merit ($FOMs$) to evaluate the proposed circuits by circuit evaluation criteria and accuracy metrics simultaneously.
\end{enumerate}

The rest of the article is arranged in 4 sections. In the second section, the background of this research are explained, and state-of-the-art is briefly introduced and reviewed. In the third section, the implementation algorithms of ECIS and ICIS1-3 full adders are introduced, and their design details are explained. The fourth section evaluates and compares the proposed circuits and state-of-the-art by circuit evaluation criteria, error analysis metrics, and image quality metrics. In subsection \ref{sec44}, the proposed approximate full adders are also evaluated by two $FOMs$ to have a comprehensive view of the reduction of circuit complexity and the reduction of accuracy in the computations of the proposed circuits. Section \ref{sec5} contains the conclusion of the article.

\section{Background} \label{sec2}
\subsection{Approximate computing} \label{sec21}
Reducing the accuracy of computations by applying AC in data-intensive error-resilient applications 
can diminish the hardware complexity of these applications.  AC has been applied to design circuits from memory to arithmetic circuits, such as adders and multipliers \cite{ref3,ref4,ref5,ref12,ref13,ref14,ref15,ref16,ref17,ref26,ref27,ref28,ref29}. Designers should focus on the compromise between reducing the accuracy of computations and 
reducing energy consumption and increasing efficiency 
in applications where AC is applied. The reasons for applying AC can be limited visual perception of humans, lack of only one acceptable answer, resistance to input noise, and error coverage and attenuation \cite{ref2,ref16}. 

To design approximate arithmetic circuits, circuit evaluation criteria such as energy consumption, performance, and area dissipation should be evaluated to ensure the improvement of circuit evaluation criteria. The evaluation of the accuracy of computations is critical from two different aspects. The first aspect is the examination of standard error evaluation criteria such as Error Rate (ER), Error Distance (ED), Mean ED (MED), and Normalized MED \cite{ref14}. By examining these criteria (see (\ref{neq1})-(\ref{neq4})), one can get a general view of the accuracy of computations by applying the designed approximate circuit in different computing structures. 
In general, the smaller the values of the error evaluation criteria, the higher the accuracy of the computation. The second aspect is to pay attention to the application in which the proposed circuits are applied. The specific output evaluation criteria of the desired application should also evaluate the designed approximate arithmetic circuits. In the application of image processing, in which the proposed approximate circuits are applied in this paper, image quality evaluation criteria such as Peak Signal to Noise Ratio (PSNR), Structural Similarity Index (SSIM), and Mean SSIM (MSSIM) should be assessed to ensure the proper performance of the approximate arithmetic circuits \cite{ref3,ref4,ref11,ref12,ref13,ref14,ref15,ref16,ref17,ref30}. The criteria for evaluating the quality of images and their importance were studied in \cite{ref11,ref31}. The greater the image quality evaluation criteria, the better the quality of the output images. The output image is acceptable and of good quality when the image created by the approximate circuit has a PSNR greater than 30 dB \cite{ref30}.

\begin{gather}
	ER= \frac{No.~of~approximate~states}{Total~no.~of~states} \label{neq1} \\
	ED= \mid A-E \mid, A:~approximate~value~\&~E:~exact~value\label{neq2} \\
	MED=\frac{\sum ED}{2^{n}}, n:No.~of~bits \label{neq3} \\
	NMED=\frac{MED}{MAX}, MAX:~maximum~output~value~of~an~exact~circuit \label{neq4}
\end{gather}

\subsection{Memristors} \label{sec22}
Emerging non-volatile memory technologies have attracted the attention of researchers in recent years, considering the importance of memory cells in all processing structures and the limitations caused by reducing the dimensions of transistors, especially in flash memories \cite{ref18}. 
Memristors can keep their previous state, and this feature makes them a suitable alternative to today's standard memory technologies. ReRAM has become a prevalent choice for replacing conventional memory technologies among researchers due to its unique features 
\cite{ref18,ref32}.

\begin{figure}
     \centering
     \begin{subfigure}[b]{0.25\textwidth}
         \centering
         \includegraphics[width=0.6\textwidth]{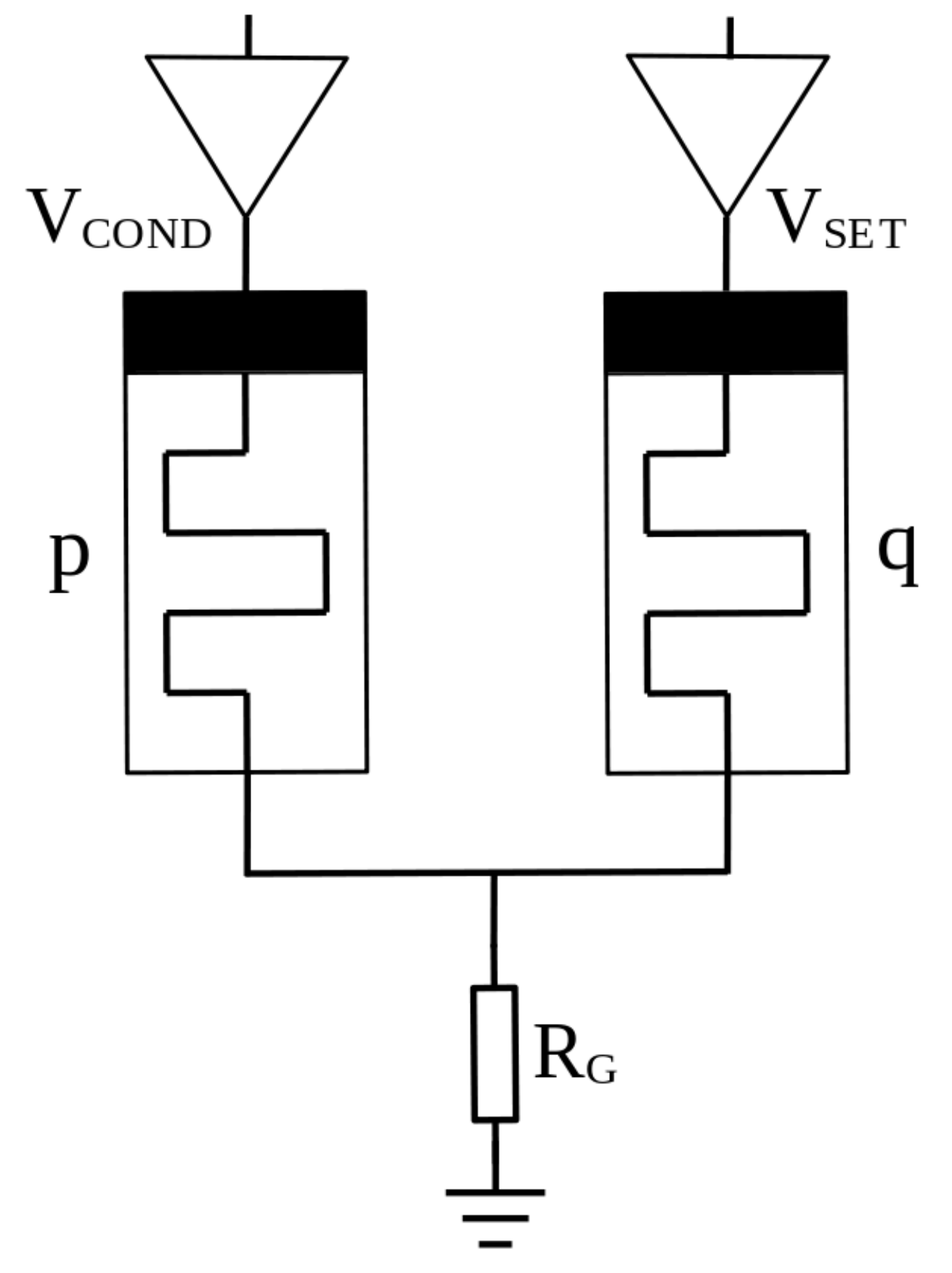}
         \caption{}
         \label{Fig1a}
     \end{subfigure}
     \begin{subfigure}[b]{0.7\textwidth}
         \centering
         \includegraphics[width=1\textwidth]{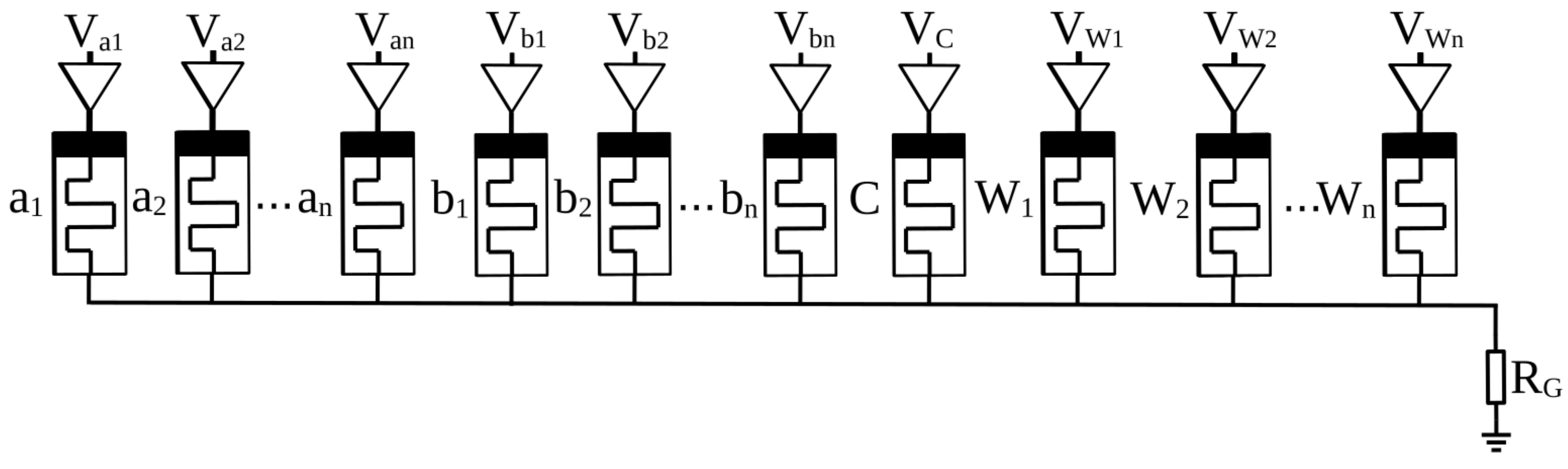}
         \caption{}
         \label{Fig1b}
     \end{subfigure}
        \caption{(a) Circuit design of a memristive IMPLY logic gate \cite{ref3}, and (b) the serial architecture of an IMPLY-based n-bit adder \cite{ref3}.}
        \label{fig1}
\end{figure}

So far, several logical design methods have been presented for circuit design applying memristors \cite{ref10,ref18,ref20,ref23,ref25}. Each of these methods has its characteristics. Statefulness is one of the essential features to consider for implementing memristor-based circuits \cite{ref3,ref18,ref24,ref33}. In the implementation of a circuit, if the input, intermediate nodes, and output logic states of a circuit are represented by resistance values, the design method of that circuit is stateful \cite{ref18}. In order to be able to perform IMP applying memristors, stateful design methods should be applied \cite{ref18,ref24}. One of the stateful and flexible memristor-based circuit design methods introduced by Hewlett-Packard (HP) is IMPLY \cite{ref3,ref13,ref18}. The IMPLY function is the primary function in the memristive IMPLY design method, done between two memristors (e.g., memristors $p$ and $q$). Applying the IMPLY function between $p$ and $q$ is represented by $p \to q$, which equals $\overline{p}+q$ \cite{ref19,ref20}.  The structure of the IMPLY logic gate is shown in Figure \ref{fig1}(a), where the memristors are connected to the ground via $R_{G}$, and its truth table is written in Table \ref{tab1}. In IMPLY logic, first, the inputs are set to HRS or LRS in memristors $p$ and $q$, and after performing the IMPLY operation, by applying voltage $V_{COND}$ to memristor $p$ and voltage $V_{SET}$ to $q$, simultaneously, the result is written in memristor $q$ \cite{ref20}. The following conditions must be met to perform IMPLY correctly \cite{ref3,ref9,ref13,ref20,ref21}:
\begin{enumerate}
	\item $V_{COND}<V_{c}<V_{SET}$, $V_{c}$ is the threshold voltage of a memristor,
	\item $R_{ON}<<R_{G}<<R_{OFF}$.
\end{enumerate}

All the logical functions (e.g., NAND, NOR, and XOR) can be implemented by applying IMPLY and FALSE (zero) functions in different numbers of computational steps. Algorithms for implementing different logic functions applying the IMPLY method are analyzed in \cite{ref19}.


\begin{table}[h]
	\centering
	\caption{The truth table of an IMPLY logic gate.}
	\scalebox{0.9}{
	\begin{tabular}{|c|c|c|}
		\hline
		$p$ & $q$ & $p~IMPLY~q$ $\equiv$ $p \to q$ \\ \hline
		0 & 0 & 1 \\ \hline
		0 & 1 & 1 \\ \hline
		1 & 0 & 0 \\ \hline
		1 & 1 & 1 \\ \hline 
	\end{tabular}}
	\label{tab1}
\end{table}

\subsection{Exact and approximate memristor-based adders} \label{sec23}
\subsubsection{IMPLY-based adders’ architectures} \label{sec231}
Four serial, parallel, semi-serial, and semi-parallel architectures were introduced to implement the algorithms of arithmetic structures (e.g., n-bit adders) applying the IMPLY method \cite{ref9,ref10,ref19,ref20,ref21}.

In serial architecture, memristors are placed next to each other in a row/column and connected to the ground by a resistor. Only a FALSE or an IMPLY operation can be executed in each computational step. The serial architecture of an n-bit adder is shown in Figure \ref{fig1}(b). Serial architecture has the simplest structure compared to other architectures. It can be well implemented in the structure of crossbar arrays \cite{ref19}. Due to the simplicity of the serial architecture and the impossibility of executing two computational steps simultaneously, the number of computational steps in this method is more than in the other architectures.

\subsubsection{IMPLY-based exact full adders} \label{sec232}
The importance of the full adder cell in all kinds of computing structures suitable for the Von-Neumann architecture and IMP is obvious. For this reason, researchers have introduced multiple IMPLY-based full adders in recent years \cite{ref9,ref19,ref20,ref21,ref34}. These full adders are designed based on all four architectures introduced in subsection \ref{sec231}. The number of required memristors, computational steps, and CMOS switches are among the most critical criteria differentiating these circuits. In Table \ref{tab2}, some of the most prominent IMPLY-based exact full adders introduced are compared with each other along with their features in the architecture applied \cite{ref9,ref19,ref20,ref21,ref34}.

\begin{table}[h]
	\centering
	\caption{Comparison between IMPLY-based exact full adders.}
	\scalebox{0.9}{
	\begin{tabular}{|c|c|c|c|}
		\hline
		Architecture & No. of & No. of & No. of \\
		& memristors & steps & switches \\ \hline
		Serial \cite{ref19} & 5 & 22 & 0 \\ \hline
		Serial \cite{ref34} & 5 & 23 & 0 \\ \hline
		Parallel \cite{ref20} & 9 & 23 & 2 \\ \hline
		Parallel \cite{ref34} & 5 & 21 & 1 \\ \hline 
		Semi-parallel \cite{ref21} & 5 & 17 & 3 \\ \hline
		Semi-serial \cite{ref9} & 8 & 12 & 12 \\ \hline
	\end{tabular}}
	\label{tab2}
\end{table}

Full adders introduced in serial architecture have less structural complexity than other full adders. However, circuits designed based on this method need more computational steps than other methods to function correctly \cite{ref9}.

\subsubsection{IMPLY-based approximate full adders} \label{sec233}
To the best of our knowledge, the first approximate full adder based on the IMPLY logic was introduced in \cite{ref13}. This full adder computes $Sum$ and $C_{out}$ outputs in 8 computational steps with only five memristors \cite{ref13}. The main problem of this approximate full adder is that the outputs are written in work memristors. As the number of approximate full adders increases in the n-bit approximate adder structure, the number of required memristors increases. The authors in \cite{ref3} solved the problem of their previously proposed approximate full adder introduced in \cite{ref13} by modifying its implementation algorithm. In \cite{ref3}, four Serial IMPLY-based Approximate Full Adders (SIAFA1-4) are introduced. SIAFA1-4 require four/five memristors (three input memristors and one/two work memristors) to implement and compute the outputs in 8, 10, 8, and 8 computational steps, respectively \cite{ref3}. Serial Approximate Full Adder using NAND gates (SAFAN) is another IMPLY-based approximate full adder that was proposed recently \cite{nref1}. This approximate full adder is designed and implemented in serial architecture like the ones proposed in \cite{ref3, ref13}. The implementation algorithm of SAFAN computes the outputs in seven computational steps by applying four memristors (three input memristors and one work memristor) \cite{nref1}. Seiler et al. proposed an approximate full adder cell (semi-serial AFA) compatible with the semi-serial architecture \cite{nref3}. The $Sum$ and $C_{out}$ outputs of this full adder are computed in six computational steps with five memristors \cite{nref3}. Error evaluation metrics of ED, MED, and NMED were calculated for the approximate full adders proposed in \cite{ref3, nref1, nref3}. The circuit and error evaluation criteria of SIAFA1-4, SAFAN, and semi-serial AFA are summarized and written in Table \ref{tab3}. SAFAN, SIAFA1-4, and semi-serial AFA were evaluated in image processing applications, too \cite{ref3, nref1, nref3}.

\begin{table}[h]
	\centering
	\caption{The circuit and error evaluation criteria of SIAFA1-4 \cite{ref3}, SAFAN \cite{nref1}, and semi-serial AFA \cite{nref3}.}
	\scalebox{0.9}{
	\begin{tabular}{|c|c|c|c|c|c|c|}
		\hline
		Approximate& No. of & No. of & No. of & ED & MED & NMED \\
		full adder & steps & memristors & switches & & & \\ \hline
		SIAFA1 \cite{ref3} & 8 & 4 & 0 & 3 & 0.375 & 0.125 \\ \hline
		SIAFA2 \cite{ref3} & 10 & 5 & 0 & 4 & 0.5 & 0.166 \\ \hline
		SIAFA3 \cite{ref3} & 8 & 4 & 0 & 3 & 0.375 & 0.125 \\ \hline
		SIAFA4 \cite{ref3} & 8 & 4 & 0 & 3 & 0.375 & 0.125 \\ \hline
		SAFAN \cite{nref1} & 7 & 4 & 0 & 3 & 0.375 & 0.125 \\ \hline
		Semi-serial AFA\cite{nref3} & 6 & 5 & 6 & 3 & 0.375 & 0.125 \\ \hline
	\end{tabular}}
	\label{tab3}
\end{table}

\section{Proposed IMPLY-based approximate full adders} \label{sec3}
In this section, the proposed circuits and their design method are described. In subsection \ref{sec31}, three approximate full adders are introduced to reduce the hardware complexity, along with an acceptable decrease in computation accuracy. In subsection \ref{sec32}, the primary purpose of proposing an approximate full adder is to avoid inexact carry propagation from LSBs to MSBs in an n-bit approximate adder structure.

\subsection{\underline{I}nexact \underline{C}arry, \underline{I}nexact \underline{S}um IMPLY-based approximate full adders (ICIS)} \label{sec31}
Three ICIS full adders (ICIS1-3) are designed in three steps and based on redesigning approximate logic from exact logic by changing the exact full adder’s truth table. It is necessary to determine the conditions of acceptability of approximate full adders’ accuracy before explaining the design steps of the proposed circuits. The maximum value of $ER_{Sum}$ and $ER_{C_{out}}$ should be $\frac{3}{8}$ and $\frac{1}{8}$ to have acceptable output results based on the error analysis metrics of state-of-the-art \cite{ref3,ref4,ref13,ref14,ref15,ref17,ref28} in error-tolerant applications, e.g., image processing. According to these ERs, the maximum acceptable value of ED for the design of the proposed approximate full adders is also considered ED of 3. Determining these conditions for designing ICIS1-3 is done because the reduction of computation’s accuracy should be limited. In addition to the error evaluation criteria, the computational steps should be significantly reduced compared to the exact full adders. The number of computational steps affects energy consumption directly.

The design steps of ICIS1-3, according to the mentioned design constraints, are as follows:

\noindent \textbf{STEP 1:} First, the truth table of the exact full adder is assessed (see Table \ref{tab4}). Then, according to the design constraints mentioned above, the $C_{out}$ of the exact full adder is inverted only in the first state ($A_{in}B_{in}C_{in}$=``000"), and $ER_{C_{out}}$ is $\frac{1}{8}$. In IMPLY logic, it is possible to invert an output in only two computational steps. In the first step, a memristor resets; in the second cycle, the IMPLY function is performed between the output memristor and the one reset in the last cycle. The $Sum$ output of these approximate full adders is assumed to be $\overline{C_{out}}$ to reduce the hardware complexity and the number of computational steps. This process is repeated for states 2 ($A_{in}B_{in}C_{in}$=``001”) to state 8 ($A_{in}B_{in}C_{in}$=``111”) of the exact full adder. Eight approximate full adders’ truth tables are designed by applying this method. We call these eight Approximate Full Adders AFA1-8. These eight approximate full adders’ truth tables can be seen next to the exact full adder’s truth table in Table \ref{tab4}. The truth tables of approximate full adders display exact outputs with \ding{51} and inexact ones with a \ding{53}. ED of each AFA specified in Table \ref{tab4}. $ER_{C_{out}}$ of these eight approximate full adders is $\frac{1}{8}$. The $ER_{Sum}$ of AFA2-7 is $\frac{3}{8}$, and the $ER_{Sum}$ of AFA1 and AFA8 equals $\frac{1}{8}$. It should be noted that the ED of all these approximate full adders equals 3. Among the AFA1-8 (see Table \ref{tab4}), the truth tables of AFA4, AFA6, and AFA7 are similar to those of SIAFA3, SIAFA1, and SIAFA4 introduced and implemented in our previous research \cite{ref3}. So, AFA1-3, 5, and 8 are considered implementation candidates according to the error analysis metrics, and the truth table of these approximate full adders is not the same as the truth tables of SIAFA1, SIAFA3, and SIAFA4 \cite{ref3}.

\begin{table}[h]
	\centering
	\caption{The truth tables of exact full adder and AFA1-8.}
	\scalebox{0.665}{
		\begin{tabular}{|c|c|c|c|c|c|c|c|c|c|c|c|c|c|c|c|c|c|c|c|c|}
			\hline
			$A_{in}$& $B_{in}$ & $C_{in}$ & Exact & Exact & AFA1 & AFA1 & AFA2 & AFA2 & AFA3 & AFA3 & AFA4 & AFA4 & AFA5 & AFA5 & AFA6 & AFA6 & AFA7 & AFA7 & AFA8 & AFA8 \\
			&  &  & $Sum$ & $C_{out}$ & $Sum$ & $C_{out}$ & $Sum$ & $C_{out}$ & $Sum$ & $C_{out}$ & $Sum$ & $C_{out}$ & $Sum$ & $C_{out}$ & $Sum$ & $C_{out}$ & $Sum$ & $C_{out}$ & $Sum$ & $C_{out}$ \\ \hline
			0 & 0 & 0 & 0 & 0 & 0 \ding{51} & 1 \ding{53} & 1 \ding{53} & 0 \ding{51} & 1 \ding{53} & 0 \ding{51} & 1 \ding{53} & 0 \ding{51} & 1 \ding{53} & 0 \ding{51} & 1 \ding{53} & 0 \ding{51} & 1 \ding{53} & 0 \ding{51} & 1 \ding{53} & 0 \ding{51} \\ \hline
			0 & 0 & 1 & 1 & 0 & 1 \ding{51} & 0 \ding{51} & 0 \ding{53} & 1 \ding{53} & 1 \ding{51} & 0 \ding{51} & 1 \ding{51} & 0 \ding{51} & 1 \ding{51} & 0 \ding{51} & 1 \ding{51} & 0 \ding{51} & 1 \ding{51} & 0 \ding{51} & 1 \ding{51} & 0 \ding{51} \\ \hline
			0 & 1 & 0 & 1 & 0 & 1 \ding{51} & 0 \ding{51} & 1 \ding{51} & 0 \ding{51} & 0 \ding{53} & 1 \ding{53} & 1 \ding{51} & 0 \ding{51} & 1 \ding{51} & 0 \ding{51} & 1 \ding{51} & 0 \ding{51} & 1 \ding{51} & 0 \ding{51} & 1 \ding{51} & 0 \ding{51} \\ \hline
			0 & 1 & 1 & 0 & 1 & 0 \ding{51} & 1 \ding{51} & 0 \ding{51} & 1 \ding{51} & 0 \ding{51} & 1 \ding{51} & 1 \ding{53} & 0 \ding{53} & 0 \ding{51} & 1 \ding{51} & 0 \ding{51} & 1 \ding{51} & 0 \ding{51} & 1 \ding{51} & 0 \ding{51} & 1 \ding{51} \\ \hline
			1 & 0 & 0 & 1 & 0 & 1 \ding{51} & 0 \ding{51} & 1 \ding{51} & 0 \ding{51} & 1 \ding{51} & 0 \ding{51} & 1 \ding{51} & 0 \ding{51} & 0 \ding{53} & 1 \ding{53} & 1 \ding{51} & 0 \ding{51} & 1 \ding{51} & 0 \ding{51} & 1 \ding{51} & 0 \ding{51} \\ \hline
			1 & 0 & 1 & 0 & 1 & 0 \ding{51} & 1 \ding{51} & 0 \ding{51} & 1 \ding{51} & 0 \ding{51} & 1 \ding{51} & 0 \ding{51} & 1 \ding{51} & 0 \ding{51} & 1 \ding{51} & 1 \ding{53} & 0 \ding{53} & 0 \ding{51} & 1 \ding{51} & 0 \ding{51} & 1 \ding{51} \\ \hline
			1 & 1 & 0 & 0 & 1 & 0 \ding{51} & 1 \ding{51} & 0 \ding{51} & 1 \ding{51} & 0 \ding{51} & 1 \ding{51} & 0 \ding{51} & 1 \ding{51} & 0 \ding{51} & 1 \ding{51} & 0 \ding{51} & 1 \ding{51} & 1 \ding{53} & 0 \ding{53} & 0 \ding{51} & 1 \ding{51} \\ \hline
			1 & 1 & 1 & 1 & 1 & 0 \ding{53} & 1 \ding{51} & 0 \ding{53} & 1 \ding{51} & 0 \ding{53} & 1 \ding{51} & 0 \ding{53} & 1 \ding{51} & 0 \ding{53} & 1 \ding{51} & 0 \ding{53} & 1 \ding{51} & 0 \ding{53} & 1 \ding{51} & 1 \ding{51} & 0 \ding{53} \\ \hline
			\multicolumn{5}{c|}{} & \multicolumn{2}{c|}{ED=3} & \multicolumn{2}{c|}{ED=3} & \multicolumn{2}{c|}{ED=3} & \multicolumn{2}{c|}{ED=3} & \multicolumn{2}{c|}{ED=3} & \multicolumn{2}{c|}{ED=3} & \multicolumn{2}{c|}{ED=3} & \multicolumn{2}{c|}{ED=3} \\ \cline{6-21}			
	\end{tabular}}
	\label{tab4}
\end{table}

In an alternative technique the truth table of eight approximate full adders is generated by inverting one bit of the exact full adder’s $Sum$ output from state $A_{in}B_{in}C_{in}$=``000” to state $A_{in}B_{in}C_{in}$=``111” similar to what was done for exact $C_{out}$. This time, $C_{out}$ equals $\overline{Sum}$, and $C_{out}$ can be calculated with only two computational steps. The truth tables of Approximate Full Adders 9-16, AFA9-16, are tabulated in Table \ref{tab5}. According to this design method, the $ER_{Sum}$ of these approximate full adders equals $\frac{1}{8}$. As in Table \ref{tab4}, the ED of AFA9-16 is written in the last row of Table \ref{tab5}. AFA9 and AFA16 have acceptable error analysis metrics (See Table \ref{tab5}) based on the conditions specified about the ER and ED of approximate full adders. According to the truth tables of these two circuits, it can be concluded that AFA9 and AFA16 are logically equivalent to AFA1 and AFA8.

Therefore, the output of this step of designing ICIS full adders consists of the truth tables of AFA1-3, AFA5, and AFA8.

\begin{table}[h]
	\centering
	\caption{The truth tables of AFA9-16.}
	\scalebox{0.665}{
		\begin{tabular}{|c|c|c|c|c|c|c|c|c|c|c|c|c|c|c|c|c|c|c|}
			\hline
			$A_{in}$& $B_{in}$ & $C_{in}$ & AFA9 & AFA9 & AFA10 & AFA10 & AFA11 & AFA11 & AFA12 & AFA12 & AFA13 & AFA13 & AFA14 & AFA14 & AFA15 & AFA15 & AFA16 & AFA16 \\
			&  &  & $Sum$ & $C_{out}$ & $Sum$ & $C_{out}$ & $Sum$ & $C_{out}$ & $Sum$ & $C_{out}$ & $Sum$ & $C_{out}$ & $Sum$ & $C_{out}$ & $Sum$ & $C_{out}$ & $Sum$ & $C_{out}$ \\ \hline
			0 & 0 & 0 & 1 \ding{53} & 0 \ding{51} & 0 \ding{51} & 1 \ding{53} & 0 \ding{51} & 1 \ding{53} & 0 \ding{51} & 1 \ding{53} & 0 \ding{51} & 1 \ding{53} & 0 \ding{51} & 1 \ding{53} & 0 \ding{51} & 1 \ding{53} & 0 \ding{51} & 1 \ding{53} \\ \hline
			0 & 0 & 1 & 1 \ding{51} & 0 \ding{51} & 0 \ding{53} & 1 \ding{53} & 1 \ding{51} & 0 \ding{51} & 1 \ding{51} & 0 \ding{51} & 1 \ding{51} & 0 \ding{51} & 1 \ding{51} & 0 \ding{51} & 1 \ding{51} & 0 \ding{51} & 1 \ding{51} & 0 \ding{51} \\ \hline
			0 & 1 & 0 & 1 \ding{51} & 0 \ding{51} & 1 \ding{51} & 0 \ding{51} & 0 \ding{53} & 1 \ding{53} & 1 \ding{51} & 0 \ding{51} & 1 \ding{51} & 0 \ding{51} & 1 \ding{51} & 0 \ding{51} & 1 \ding{51} & 0 \ding{51} & 1 \ding{51} & 0 \ding{51} \\ \hline
			0 & 1 & 1 & 0 \ding{51} & 1 \ding{51} & 0 \ding{51} & 1 \ding{51} & 0 \ding{51} & 1 \ding{51} & 1 \ding{53} & 0 \ding{53} & 0 \ding{51} & 1 \ding{51} & 0 \ding{51} & 1 \ding{51} & 0 \ding{51} & 1 \ding{51} & 0 \ding{51} & 1 \ding{51} \\ \hline
			1 & 0 & 0 & 1 \ding{51} & 0 \ding{51} & 1 \ding{51} & 0 \ding{51} & 1 \ding{51} & 0 \ding{51} & 1 \ding{51} & 0 \ding{51} & 0 \ding{53} & 1 \ding{53} & 1 \ding{51} & 0 \ding{51} & 1 \ding{51} & 0 \ding{51} & 1 \ding{51} & 0 \ding{51} \\ \hline
			1 & 0 & 1 & 0 \ding{51} & 1 \ding{51} & 0 \ding{51} & 1 \ding{51} & 0 \ding{51} & 1 \ding{51} & 0 \ding{51} & 1 \ding{51} & 0 \ding{51} & 1 \ding{51} & 1 \ding{53} & 0 \ding{53} & 0 \ding{51} & 1 \ding{51} & 0 \ding{51} & 1 \ding{51} \\ \hline
			1 & 1 & 0 & 0 \ding{51} & 1 \ding{51} & 0 \ding{51} & 1 \ding{51} & 0 \ding{51} & 1 \ding{51} & 0 \ding{51} & 1 \ding{51} & 0 \ding{51} & 1 \ding{51} & 0 \ding{51} & 1 \ding{51} & 1 \ding{53} & 0 \ding{53} & 0 \ding{51} & 1 \ding{51} \\ \hline
			1 & 1 & 1 & 1 \ding{51} & 0 \ding{53} & 1 \ding{51} & 0 \ding{53} & 1 \ding{51} & 0 \ding{53} & 1 \ding{51} & 0 \ding{53} & 1 \ding{51} & 0 \ding{53} & 1 \ding{51} & 0 \ding{53} & 1 \ding{51} & 0 \ding{53} & 0 \ding{53} & 1 \ding{51} \\ \hline
			\multicolumn{3}{c|}{} & \multicolumn{2}{c|}{ED=3} & \multicolumn{2}{c|}{ED=5} & \multicolumn{2}{c|}{ED=5} & \multicolumn{2}{c|}{ED=5} & \multicolumn{2}{c|}{ED=5} & \multicolumn{2}{c|}{ED=5} & \multicolumn{2}{c|}{ED=5} & \multicolumn{2}{c|}{ED=3} \\ \cline{4-19}			
	\end{tabular}}
	\label{tab5}
\end{table}

\noindent \textbf{STEP 2:} After determining the truth table of the acceptable circuits from STEP 1, the Boolean logic function of the outputs of each circuit must be determined. Based on these functions, an algorithm can be provided for implementing these circuits with IMPLY logic by applying the IMPLY and FALSE functions to memristors. The Karnaugh map is one of the methods that can be applied to simplify the Boolean output functions of combinational circuits. In this step, the Karnaugh map is applied to simplify and design the output functions of each approximate full adder.

AFA1-3, AFA5, and AFA8 are selected approximate circuits from the previous step. Each of these approximate full adders has two approximate outputs, and for each of the outputs, Boolean functions are specified by applying the Karnaugh map. Logic functions of the outputs of AFA1-3, AFA5, and AFA8 are written in (\ref{eq1})-(\ref{eq10}), respectively. The output of this step of designing the ICIS full adders is the Boolean logic functions of (\ref{eq1})-(\ref{eq10}).

\begin{gather}
	Sum_{AFA1}=\overline{A_{in}} \cdot (B_{in} \oplus C_{in}) + \overline{C_{in}} \cdot (A_{in} \oplus B_{in}) \label{eq1} \\
	C_{out_{AFA1}}=\overline{Sum_{AFA1}} \label{eq2}
\end{gather}

\begin{gather}
	Sum_{AFA2}=\overline{C_{in}} \cdot (\overline{B_{in} \cdot A_{in}}) \label{eq3} \\
	C_{out_{AFA2}}=\overline{Sum_{AFA2}} \label{eq4}
\end{gather}

\begin{gather}
	Sum_{AFA3}=\overline{B_{in}} \cdot (\overline{A_{in} \cdot C_{in}}) \label{eq5} \\
	C_{out_{AFA3}}=\overline{Sum_{AFA3}} \label{eq6}
\end{gather}

\begin{gather}
	Sum_{AFA5}=\overline{A_{in}} \cdot (\overline{B_{in} \cdot C_{in}}) \label{eq7} \\
	C_{out_{AFA5}}=\overline{Sum_{AFA5}} \label{eq8}
\end{gather}

\begin{gather}
	\begin{split}
			Sum_{AFA8}=\overline{A_{in}} \cdot (\overline{B_{in}\cdot C_{in}})+\overline{B_{in}} \cdot \overline{C_{in}} +A_{in}\cdot B_{in} \cdot C_{in} \label{eq9}
	\end{split} \\
	C_{out_{AFA8}}=\overline{Sum_{AFA8}}. \label{eq10}
\end{gather}

\noindent \textbf{STEP 3:} The logic functions in (\ref{eq1})-(\ref{eq10}) are designed based on Boolean logic gates. Applying these logic gates in the design of IMPLY-based circuits is impossible. (\ref{eq1})-(\ref{eq10}) should be rewritten so that it is possible to implement by applying IMPLY and FALSE functions. Implementing all Boolean logic functions applying IMPLY and FALSE functions is possible. The implementation of binary logic gates required by (\ref{eq1})-(\ref{eq10}) can be investigated and studied in \cite{ref19}. (\ref{eq1})-(\ref{eq10}) are rewritten, applying IMPLY and FALSE functions in (\ref{eq11})-(\ref{eq20}).

\begin{gather}
	\begin{split}
		Sum_{AFA1}=[B_{in}\to(\overline{C_{in}}\to A_{in})] \to [(((A_{in} \to C_{in})	\to(\overline{\overline{A_{in}} \to \overline{C_{in}}}))\to B_{in}) \to 0] \label{eq11}
	\end{split} \\
	C_{out_{AFA1}}=Sum_{AFA1} \to 0 \label{eq12}
\end{gather}

\begin{gather}
	C_{out_{AFA2}}=(B_{in}\to(A_{in}\to 0))\to C_{in} \label{eq13} \\
	Sum_{AFA2}=C_{out_{AFA2}} \to 0 \label{eq14}
\end{gather}

\begin{gather}
	C_{out_{AFA3}}=(C_{in}\to(A_{in}\to 0))\to B_{in} \label{eq15} \\
	Sum_{AFA3}=C_{out_{AFA3}} \to 0 \label{eq16}
\end{gather}

\begin{gather}
	C_{out_{AFA5}}=(C_{in}\to(B_{in}\to 0))\to A_{in} \label{eq17} \\
	Sum_{AFA5}=C_{out_{AFA5}} \to 0 \label{eq18}
\end{gather}

\begin{gather}
	\begin{split}
		Sum_{AFA8}=[\overline{B_{in}}\to C_{in}]\to[(((B_{in}\to \overline{C_{in}})\to A_{in}) \to (\overline{B_{in}\to (C_{in}\to \overline{A_{in}})}))] \label{eq19}
	\end{split} \\
	C_{out_{AFA8}}=Sum_{AFA8} \to 0 \label{eq20}
\end{gather}

The method of computing the output of $C_{out_{AFA2}}$ is described as an example to better understand how (\ref{eq11})-(\ref{eq20}) are written. To generate the output of $C_{out_{AFA2}}$, direct implementation of the functions of NAND and AND logic gates based on IMPLY and FALSE functions and overlapping of the steps related to the FALSE functions in the implementation of these two logic gates ((\ref{neq21}) and (\ref{neq22})) has been applied. As can be seen in (\ref{neq23}), it is possible to overlap the two FALSE functions and eliminate their consecutive steps. Therefore, using direct implementation of Boolean logic gates using the IMPLY and FALSE functions and overlapping of the steps is the general method used to compute (\ref{eq11})-(\ref{eq20}). The method of implementing basic logic gates using the IMPLY and FALSE functions is explained in \cite{ref19}.

\begin{gather}
	\overline{A_{in} \cdot B_{in}} \equiv (B_{in} \to (A_{in} \to 0)):\alpha \label{neq21} \\
	\overline{C_{in}} \cdot (\overline{A_{in} \cdot B_{in}}) \equiv \overline{C_{in}} \cdot \alpha \equiv (\alpha \to ((C_{in} \to 0) \to 0)) \to 0 \equiv (\alpha \to C_{in}) \to 0  \equiv ((B_{in} \to (A_{in} \to 0)) \to C_{in}) \to 0\label{neq22} \\
		C_{out_{AFA2}}=\overline{\overline{C_{in}} \cdot (\overline{B_{in} \cdot A_{in}})} \equiv (\overline{C_{in}} \cdot (\overline{B_{in} \cdot A_{in}})) \to 0 \equiv (((B_{in} \to (A_{in} \to 0)) \to C_{in}) \to 0) \to 0 \nonumber \\
		 \equiv (B_{in}\to(A_{in}\to 0))\to C_{in}  \label{neq23}
\end{gather}

\begin{figure}[h]
	\centering
	\includegraphics[scale=0.28]{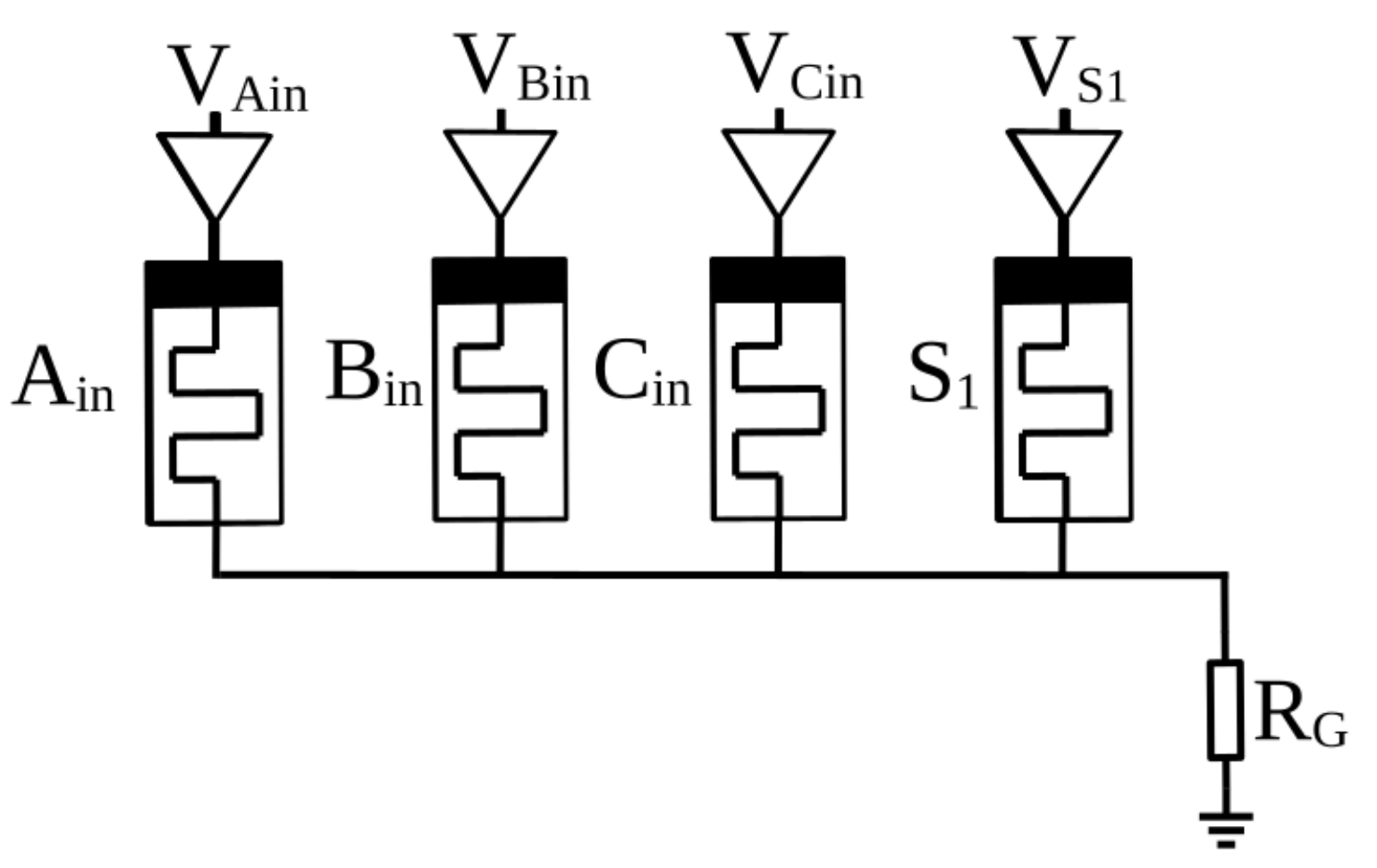}
	\caption{Schematic representation of ICIS1, ICIS2, and ICIS3.}
	\label{fig2n}
\end{figure}

Now, (\ref{eq11})-(\ref{eq20}) should be implemented serially based on the structure of Figure \ref{fig2n}  by applying a few memristors, three input memristors ($A_{in}$, $B_{in}$, and $C_{in}$), and a maximum of two work memristors ($S_{1}$ and $S_{2}$). The serial implementation algorithms of AFA2, AFA3, and AFA5 are written in Tables \ref{tab6}-\ref{tab8}, respectively. These three approximate full adders can be implemented in 6 computational steps according to the presented algorithms and by applying four memristors, including three input memristors and a work memristor.

\begin{table}[h]
	\centering
	\caption{AFA2's IMPLY-based implementation algorithm (ICIS1).}
	\scalebox{0.9}{
		\begin{tabular}{|c|c|c|}
			\hline
			Step & Operation & Equivalent logic \\ \hline
			1 & $S_{1}=0$ & $FALSE(S_{1})$ \\ \hline
			2 & $A_{in} \to S_{1}=S_{1}^{'}$ & $NOT(A_{in})$\\ \hline
			3 & $B_{in} \to S_{1}^{'}=S_{1}^{''}$ & $B_{in} \to NOT(A_{in})$ \\ \hline
			4 & $S_{1}^{''} \to C_{in}=C_{in}^{'}$ & $C_{out}=(B_{in}\to(A_{in}\to 0))\to C_{in}$ \\ \hline
			5 & $A_{in}=0$ & $FALSE(A_{in})$ \\ \hline
			6 & $C_{in}^{'} \to A_{in}=A_{in}^{'}$ & $Sum=\overline{C_{out}}$ \\ \hline
	\end{tabular}}
	\label{tab6}
\end{table}

\begin{table}[h]
	\centering
	\caption{AFA3's IMPLY-based implementation algorithm (ICIS2).}
	\scalebox{0.9}{
		\begin{tabular}{|c|c|c|}
			\hline
			Step & Operation & Equivalent logic \\ \hline
			1 & $S_{1}=0$ & $FALSE(S_{1})$ \\ \hline
			2 & $A_{in} \to S_{1}=S_{1}^{'}$ & $NOT(A_{in})$\\ \hline
			3 & $C_{in} \to S_{1}^{'}=S_{1}^{''}$ & $C_{in} \to NOT(A_{in})$ \\ \hline
			4 & $S_{1}^{''} \to B_{in}=B_{in}^{'}$ & $C_{out}=(C_{in}\to(A_{in}\to 0))\to B_{in}$ \\ \hline
			5 & $A_{in}=0$ & $FALSE(A_{in})$ \\ \hline
			6 & $B_{in}^{'} \to A_{in}=A_{in}^{'}$ & $Sum=\overline{C_{out}}$ \\ \hline
	\end{tabular}}
	\label{tab7}
\end{table}

\begin{table}[!h]
	\centering
	\caption{AFA5's IMPLY-based implementation algorithm (ICIS3).}
	\scalebox{0.9}{
		\begin{tabular}{|c|c|c|}
			\hline
			Step & Operation & Equivalent logic \\ \hline
			1 & $S_{1}=0$ & $FALSE(S_{1})$ \\ \hline
			2 & $B_{in} \to S_{1}=S_{1}^{'}$ & $NOT(B_{in})$\\ \hline
			3 & $C_{in} \to S_{1}^{'}=S_{1}^{''}$ & $C_{in} \to NOT(B_{in})$ \\ \hline
			4 & $S_{1}^{''} \to A_{in}=A_{in}^{'}$ & $C_{out}=(C_{in}\to(B_{in}\to 0))\to A_{in}$ \\ \hline
			5 & $B_{in}=0$ & $FALSE(B_{in})$ \\ \hline
			6 & $A_{in}^{'} \to B_{in}=B_{in}^{'}$ & $Sum=\overline{C_{out}}$ \\ \hline
	\end{tabular}}
	\label{tab8}
\end{table}

The implementation algorithm of AFA1 and AFA8 is written serially by applying IMPLY and FALSE functions according to the output functions of these two approximate full adders, (\ref{eq11}), (\ref{eq12}), (\ref{eq19}), and (\ref{eq20}). These approximate full adders are removed from the set of ICIS full adders candidates. The main reason for this decision is that, according to the complexity of these two circuits and their implementation algorithms, at least five memristors are needed (three input memristors and at least two work memristors) to implement them in 17 computational steps serially. Comparing the computational steps required by these two circuits and their ED with AFA2, AFA3, and AFA5, it can be concluded that the accuracy of computations applying these two circuits will not increase significantly. In contrast, the number of computational steps increases by 65\% compared to AFA2, AFA3, and AFA5. So, there would be no justification for presenting these circuits as ICIS approximate full adders.

AFA2, AFA3, and AFA5 are called ICIS1, ICIS2, and ICIS3, respectively, in the rest of the article and Tables \ref{tab6}-\ref{tab8}.

\subsection{\underline{E}xact \underline{C}arry, \underline{I}nexact \underline{S}um IMPLY-based approximate full adder (ECIS)} \label{sec32}
The primary purpose of designing ECIS is to improve its error evaluation criteria compared to ICIS1-3. Applying this full adder avoids error propagation from the LSBs to MSBs in n-bit approximate adders. The ECIS full adder is designed in three steps, like ICIS1-3. These three steps are:

\noindent \textbf{STEP 1:} Designing an approximate full adder in which its $C_{out}$ is computed accurately is a solution to prevent error propagation from the LSBs. In ECIS, the $C_{out}$ is exact, and the $Sum$ is considered $\overline{C_{out}}$. The truth table of ECIS can be seen in Table \ref{tab9}. According to Table \ref{tab9}, this approximate full adder’s $ER_{Sum}$ is $\frac{2}{8}$, and $C_{out}$ is calculated precisely ($ER_{C_{out}}$ is 0). The ED of ECIS is 2. As in Tables \ref{tab4} and \ref{tab5}, the exact states are marked with a \ding{51}, and the inexact states are labeled with a \ding{53}. By examining the error evaluation criteria, it can be concluded that the accuracy of ECIS computations will be higher than ICIS1-3.

\begin{table}[h]
	\centering
	\caption{The truth tables of exact full adder and ``ECIS" \cite{ref27}.}
	\scalebox{0.9}{
		\begin{tabular}{|c|c|c|c|c|c|c|}
			\hline
			$A_{in}$& $B_{in}$ & $C_{in}$ & Exact & Exact & ECIS & ECIS \\
			&  &  & $Sum$ & $C_{out}$ & $Sum$ & $C_{out}$  \\ \hline
			0 & 0 & 0 & 0 & 0 & 1 \ding{53} & 0 \ding{51}  \\ \hline
			0 & 0 & 1 & 1 & 0 & 1 \ding{51} & 0 \ding{51}  \\ \hline
			0 & 1 & 0 & 1 & 0 & 1 \ding{51} & 0 \ding{51}  \\ \hline
			0 & 1 & 1 & 0 & 1 & 0 \ding{51} & 1 \ding{51}  \\ \hline
			1 & 0 & 0 & 1 & 0 & 1 \ding{51} & 0 \ding{51}  \\ \hline
			1 & 0 & 1 & 0 & 1 & 0 \ding{51} & 1 \ding{51}  \\ \hline
			1 & 1 & 0 & 0 & 1 & 0 \ding{51} & 1 \ding{51}  \\ \hline
			1 & 1 & 1 & 1 & 1 & 0 \ding{53} & 1 \ding{51}  \\ \hline
			\multicolumn{5}{c|}{} & \multicolumn{2}{c|}{ED=2}  \\ \cline{6-7}			
	\end{tabular}}
	\label{tab9}
\end{table}

\noindent \textbf{STEP 2:} In this step, the logic functions of the outputs of ECIS are written in (\ref{eq21}) and (\ref{eq22}) according to Table \ref{tab9} and by applying the Karnaugh map. (\ref{eq21}) and (\ref{eq22}), as outputs of this step, should be implemented serially by applying IMPLY and FALSE functions.

\begin{gather}
	Sum_{ECIS}=(\overline{B_{in}} \cdot (\overline{A_{in} \cdot C_{in}})) + (\overline{A_{in} + C_{in}}) \label{eq21} \\
	C_{out_{ECIS}}=\overline{Sum_{ECIS}} \label{eq22}
\end{gather}

\noindent \textbf{STEP 3:} A serial algorithm and five memristors (three input memristors and a maximum of two work memristors) are needed to implement an IMPLY-based ECIS. First, we implement (\ref{eq21}) and (\ref{eq22}) by applying IMPLY and FALSE functions. The $Sum$ output of the ECIS can be implemented by IMPLY logic as
\begin{gather}
	\begin{split}
		[((C_{in}\to 0)\to A_{in}) \to (((C_{in}\to (A_{in}\to 0)) \to B_{in}) \to 0)] \label{eq23}
	\end{split}
\end{gather}

and its $C_{out}$ equals $\overline{Sum}$.
\begin{gather}
	C_{out_{ECIS}}=Sum_{ECIS} \to 0 \label{eq24}
\end{gather}

\begin{figure}[h]
	\centering
	\includegraphics[scale=0.275]{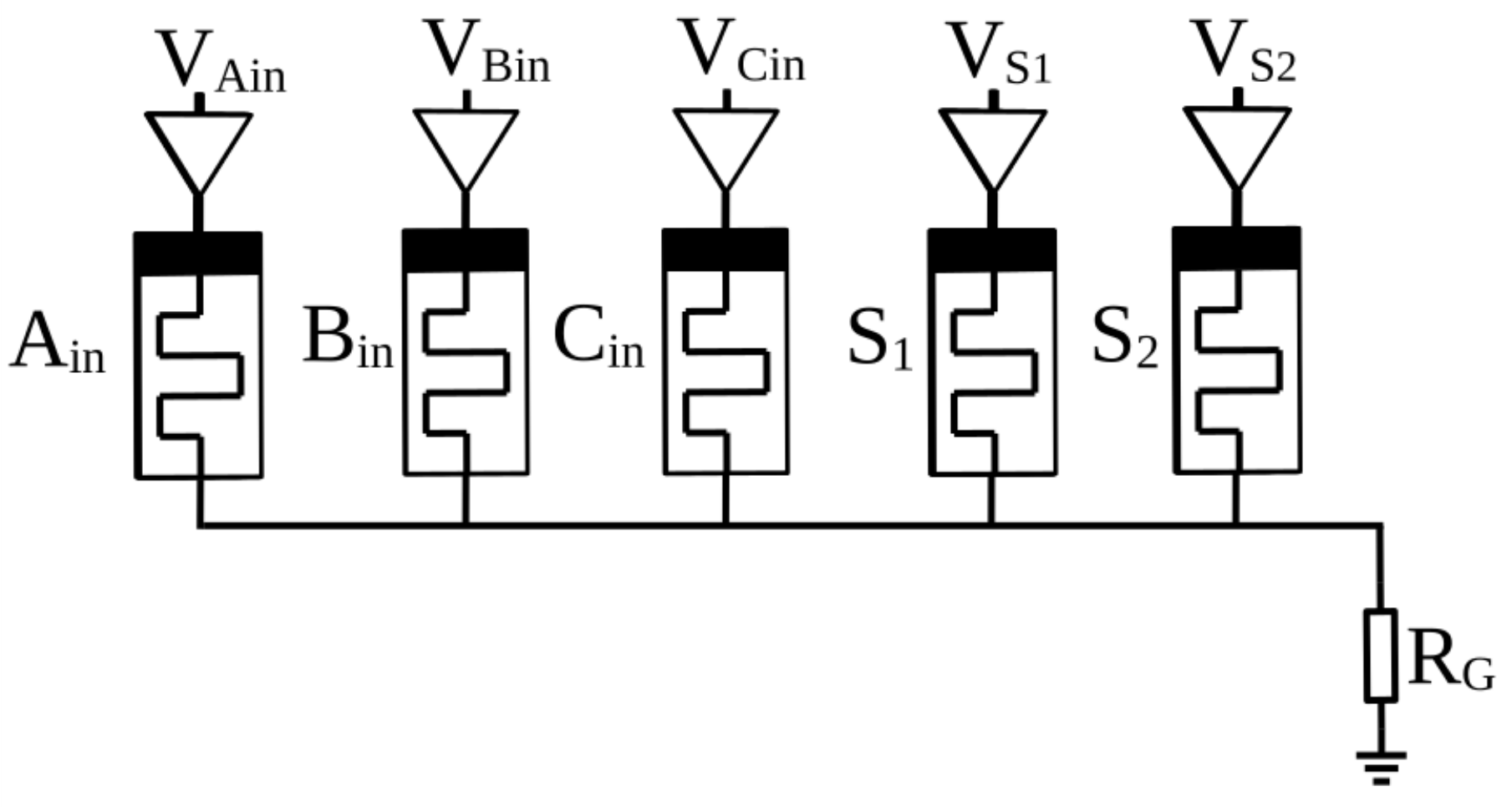}
	\caption{Schematic representation of ECIS.}
	\label{fig3n}
\end{figure}

In Table \ref{tab10}, the serial implementation algorithm of ECIS is written in 12 steps by applying five memristors ($A_{in}$, $B_{in}$, $C_{in}$, $S_{1}$, and $S_{2}$). In this algorithm, $Sum$ is calculated in the tenth step, and $C_{out}$ is calculated in the twelfth step. The circuit-level implementation of ECIS is depicted in Figure \ref{fig3n}.

\begin{table}[h]
	\centering
	\caption{IMPLY-based implementation algorithm of ECIS.}
	\scalebox{0.9}{
		\begin{tabular}{|c|c|c|}
			\hline
			Step & Operation & Equivalent logic \\ \hline
			1 & $S_{1}=0$ & $FALSE(S_{1})$ \\ \hline
			2 & $S_{2}=0$ & $FALSE(S_{2})$ \\ \hline
			3 & $A_{in} \to S_{1}=S_{1}^{'}$ & $NOT(A_{in})$ \\ \hline
			4 & $C_{in} \to S_{2}=S_{2}^{'}$ & $NOT(C_{in})$ \\ \hline
			5 & ${S_{2}^{'}} \to A_{in}= A_{in}^{'}$ & $NOT(C_{in}) \to A_{in}$ \\ \hline
			6 & $C_{in} \to S_{1}^{'}=S_{1}^{''}$ & $C_{in} \to NOT(A_{in})$ \\ \hline
			7 & $S_{1}^{''} \to B_{in}=B_{in}^{'}$ & $(C_{in} \to NOT(A_{in})) \to B_{in}$ \\ \hline
			8 & $C_{in}=0$ & $FALSE(C_{in})$ \\ \hline
			9 & $B_{in}^{'}\to C_{in}=C_{in}^{'}$ & $NOT((C_{in} \to NOT(A_{in})) \to B_{in})$ \\ \hline
			10 & $A_{in}^{'} \to C_{in}^{'}=C_{in}^{''}$ & $Sum=[((C_{in}\to 0)\to A_{in}) \to $ \\
			& & $(((C_{in}\to (A_{in}\to 0)) \to B_{in}) \to 0)]$\\ \hline
			11 & $B_{in}=0$ & $FALSE(B_{in})$  \\ \hline
			12 & $C_{in}^{''} \to B_{in}=B_{in}^{'}$ & $C_{out}=\overline{Sum}$ \\ \hline
	\end{tabular}}
	\label{tab10}
\end{table}

\subsection{Summary of the proposed approximate full adders} \label{sec33}
This subsection summarizes the properties of the proposed approximate full adders introduced in subsections \ref{sec31} and \ref{sec32}. The main features of ICIS1-3 and ECIS, including error analysis metrics, number of computational steps, and required number of memristors, are presented in Table \ref{tab11}.

\begin{table}[h]
	\centering
	\caption{The circuit and error evaluation criteria of ICIS1-3 and ECIS.}
	\scalebox{0.9}{
		\begin{tabular}{|c|c|c|c|c|c|}
			\hline
			Approximate& No. of & No. of & ED & MED & NMED \\
			full adder & steps & memristors & & & \\ \hline
			ICIS1 & 6 & 4 & 3 & 0.375 & 0.125 \\ \hline
			ICIS2 & 6 & 4 & 3 & 0.375 & 0.125 \\ \hline
			ICIS3 & 6 & 4 & 3 & 0.375 & 0.125 \\ \hline
			ECIS & 12 & 5 & 2 & 0.25 & 0.0833 \\ \hline
	\end{tabular}}
	\label{tab11}
\end{table}

\section{Simulation results and comparison} \label{sec4}
In this section, the proposed approximate full adders and state-of-the-art \cite{ref3,ref19,ref34, nref1} are compared with each other by circuit evaluation criteria (subsection \ref{sec41}), error analysis metrics (subsection \ref{sec42}), and image quality metrics (subsection \ref{sec43}). Then, in subsection \ref{sec44}, the results of these three simulations are compared with each other by two $FOMs$.

\subsection{Circuit-level simulation and analysis} \label{sec41}
The hardware complexity of the proposed circuits is reduced by applying approximate computing. Approximate full adders’ circuit-level simulation results (energy consumption, computational steps, and estimation of area dissipation) compared to the exact \cite{ref19,ref34} and approximate full adders \cite{ref3,nref1} are critical in determining the improvement of circuit evaluation criteria. Accordingly, it is possible to calculate the improvement of circuit evaluation criteria and apply the simulation results to analyze the application of the proposed circuits in larger computing structures. LTSPICE simulator and Voltage ThrEshold Adaptive Memristor (VTEAM) model are applied for circuit-level simulation of the proposed circuits, and state-of-the-art \cite{ref3,ref19,ref34,nref1}. The semi-serial architecture's compatibility with the crossbar array is moderate \cite{nref4}; hence, the proposed circuits designed based on serial architecture are not comparable with the semi-serial AFA \cite{nref3}. In Table \ref{tab12}, the parameters of the applied memristor model and the required parameters for simulating the IMPLY function are written.

\begin{table}[t]
	\centering
	\caption{Setup values of VTEAM model and IMPLY logic \cite{ref3,ref9,nref1}.}
	\scalebox{0.9}{
		\begin{tabular}{|c|c|c|c|c|c|}
			\hline
			Parameter & Value & Parameter & Value & Parameter & Value \\ \hline
			$v_{off}$ & 0.7 V & $v_{on}$ & -10 mV & $\alpha_{off}$ & 3 \\ \hline
			$\alpha_{on}$ & 3 & $R_{off}$ & 1 M$\Omega$ & $R_{on}$ & 10 k$\Omega$ \\ \hline
			$k_{on}$ & -0.5 $\frac{nm}{s}$ & $k_{off}$ & 1 $\frac{cm}{s}$ & $w_{off}$ & 0 nm \\ \hline
			$w_{on}$ & 3 nm & $w_{C}$ & 107 $pm$ & $a_{off}$ & 3 $nm$ \\ \hline
			$a_{on}$ & 0 $nm$ & $v_{set}$ & 1 V & $v_{reset}$ & 1 V \\ \hline
			$v_{cond}$ & 900 mV & $R_{G}$ & 40 K$\Omega$ & $t_{pulse}$ & 30 $\mu$s \\ \hline
	\end{tabular}}
	\label{tab12}
\end{table}

\begin{figure}[h!]
	\centering
	\includegraphics[scale=0.215]{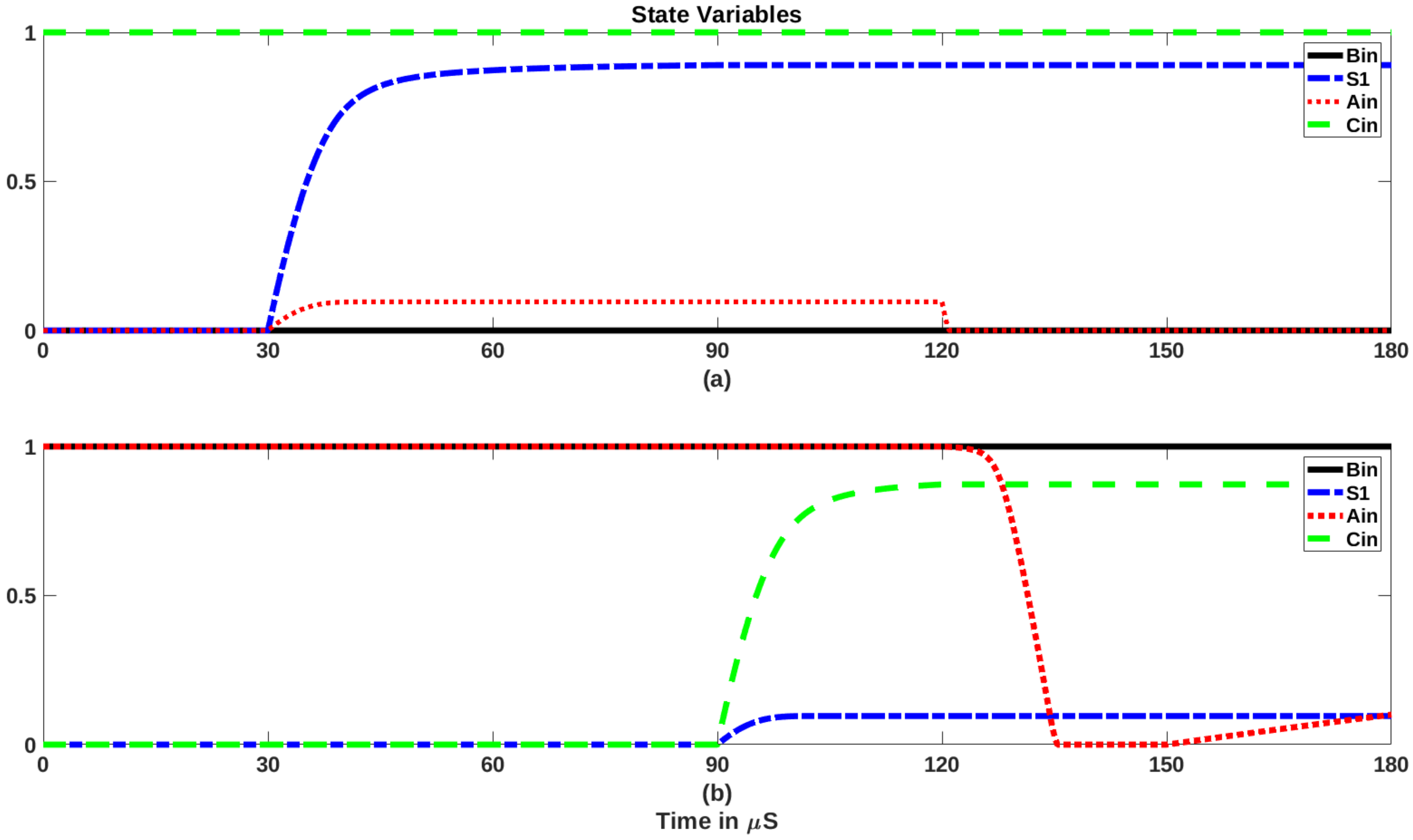}
	\caption{ICIS1's waveforms: (a) $A_{in}B_{in}C_{in}$=$``001"$, and (b) $A_{in}B_{in}C_{in}$=$``110"$.}
	\label{fig2}
\end{figure}

\begin{figure}[h!]
	\centering
	\includegraphics[scale=0.215]{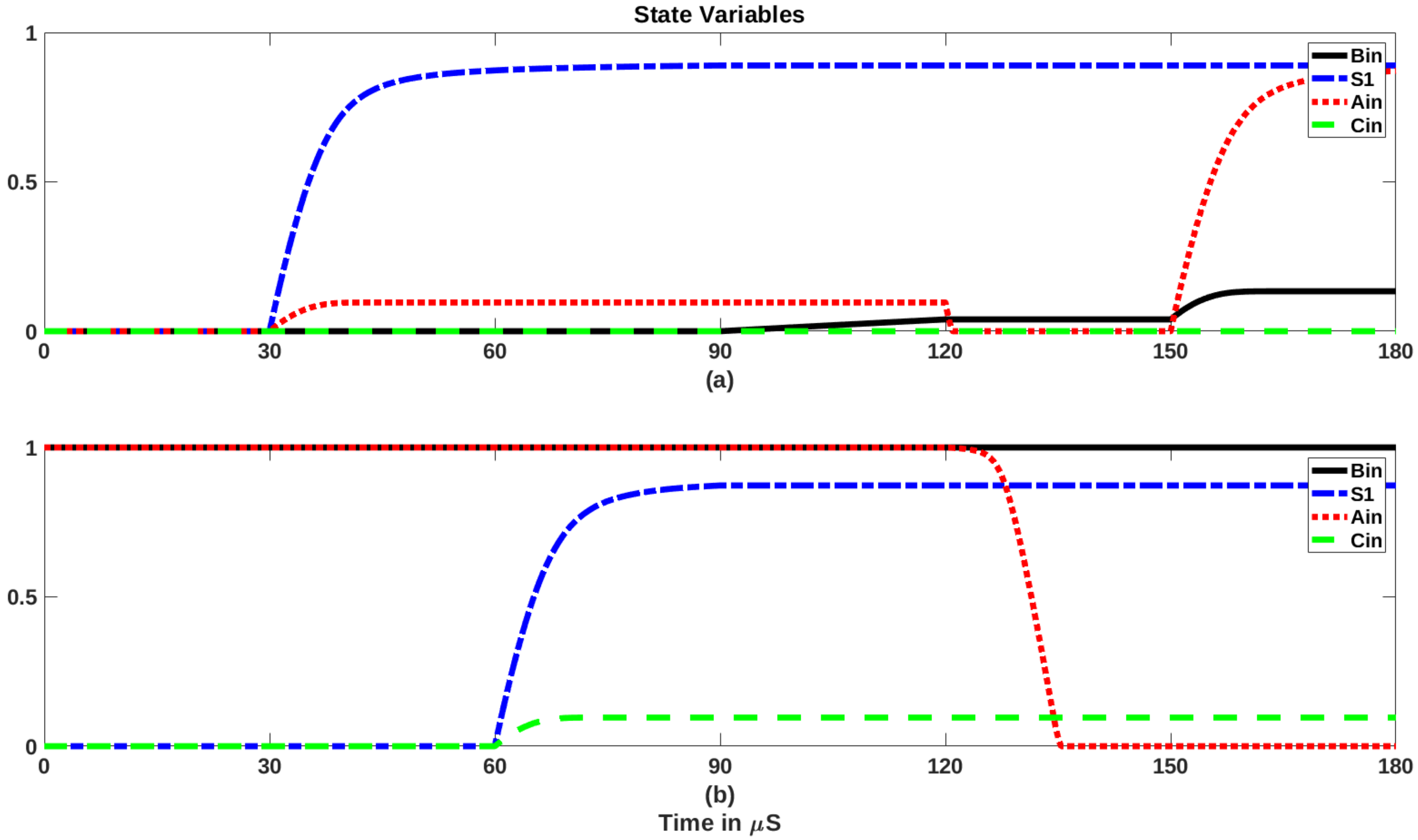}
	\caption{ICIS2's waveforms: (a) $A_{in}B_{in}C_{in}$=$``000"$, and (b) $A_{in}B_{in}C_{in}$=$``110"$.}
	\label{fig3}
\end{figure}

\begin{figure}[h!]
	\centering
	\includegraphics[scale=0.215]{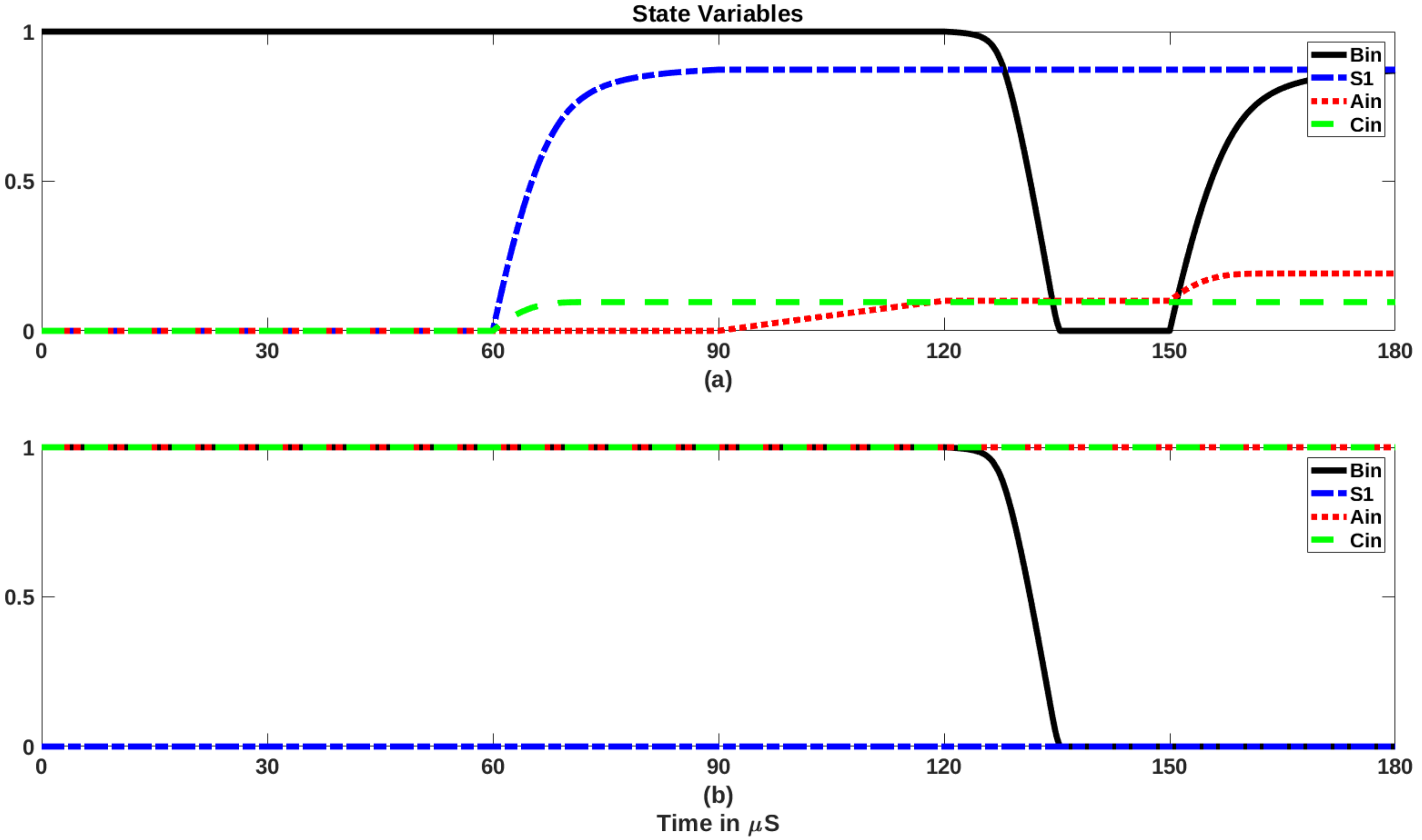}
	\caption{ICIS3's waveforms: (a) $A_{in}B_{in}C_{in}$=$``010"$, and (b) $A_{in}B_{in}C_{in}$=$``111"$.}
	\label{fig4}
\end{figure}

\begin{figure}[h!]
	\centering
	\includegraphics[scale=0.215]{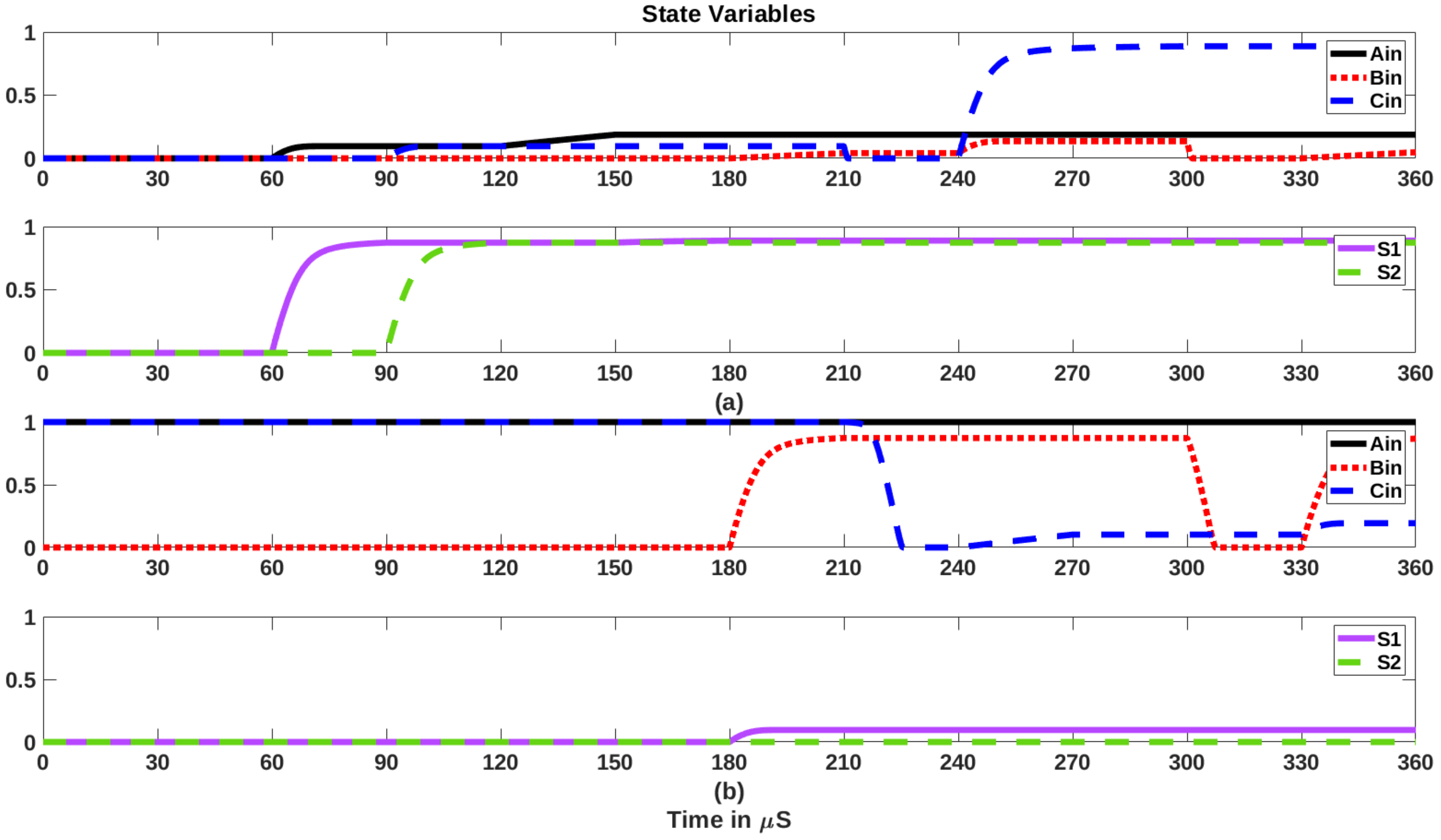}
	\caption{ECIS's waveforms: (a) $A_{in}B_{in}C_{in}$=$``000"$, and (b) $A_{in}B_{in}C_{in}$=$``101"$.}
	\label{fig5}
\end{figure}

The proposed circuits are simulated by applying the parameters reported in Table \ref{tab12} and all the input patterns considered. According to the simulation results, the proposed algorithms for implementing ICIS1-3 and ECIS led to correct results in all cases. Two output waveforms of each of the circuits, including inputs that lead to approximate and exact outputs, as showcased, are shown in Figures \ref{fig2}-\ref{fig5}. In Figures \ref{fig2}-\ref{fig5}, logic ‘1’ corresponds to LRS, and logic ‘0’ corresponds to HRS.

ICIS1-3 calculate the $Sum$ and $C_{out}$ outputs in 6 computational steps. Each computational step is considered 30 $\mu$s \cite{ref3,ref9,nref1,nref5,nref6}. The $C_{out}$ and $Sum$ outputs of each of ICIS1-3 are computed in the fourth (90-120 $\mu$s) and sixth (150-180 $\mu$s) computational steps. Figures \ref{fig2}-\ref{fig4} show the output waveforms of the ICIS1-3. In each of Figures \ref{fig2}-\ref{fig4}, two different inputs and the corresponding outputs of each input can be seen.

The output waveforms of the ECIS for two input states can be seen in Figure \ref{fig5}. The inputs of $A_{in}B_{in}C_{in}$=``000” and $A_{in}B_{in}C_{in}$=``101” applied to ECIS, its implementation algorithm ran, and the output waveforms depicted in Figure \ref{fig5}. In this circuit, $Sum$ is stored in memristor ‘$C_{in}$’ in the tenth step (270-300 $\mu$s), and $C_{out}$ is stored in memristor ‘$B_{in}$’ in the twelfth step (330-360 $\mu$s).

\begin{figure}[h]
	\centering
	\includegraphics[scale=0.22]{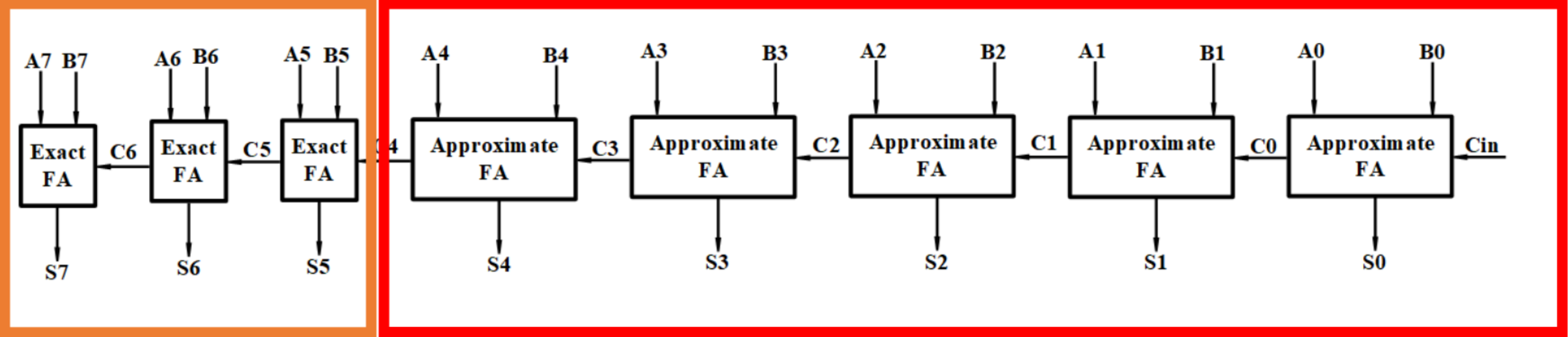}
	\caption{The third scenario's RCA structure \cite{ref4}.}
	\label{fig6}
\end{figure}

Approximate full adders can be applied to n-bit approximate adders in different ways. Generally,  approximate full adders are placed in the LSBs, and the exact full adders are applied in the MSBs to control the accuracy of computations. An approximate 8-bit Ripple Carry Adder (RCA) structure is shown in Figure \ref{fig6}. In this structure, 5 LSBs are computed by approximate full adders, and exact full adders calculate 3 MSBs. Three scenarios are considered to report and compare the results of different simulations in this article. The structure in which the mentioned scenarios are examined is the 8-bit approximate RCA. Three, four, and five LSBs of the 8-bit RCA structure are calculated by the approximate full adders in the first, second, and third scenarios, respectively. In the approximate RCA structures of scenarios 1-3, the MSBs (5, 4, and 3, respectively) are calculated by the exact full adders.

\begin{table}[h]
	\centering
	\caption{Comparison of the number of computational steps and memristors of ICIS1-3, ECIS, SIAFA1-4 \cite{ref3}, SAFAN \cite{nref1}, and exact full adders in \cite{ref19} and \cite{ref34}.}
	\scalebox{0.75}{
		\begin{tabular}{|c|c|c|c|c|}
			\hline
			Serial & \multicolumn{2}{c|}{No. of} & \multicolumn{2}{c|}{No. of} \\
			full adder & \multicolumn{2}{c|}{steps} & \multicolumn{2}{c|}{memristors} \\ \cline{2-5}
			(exact and approximate) & n & n=8-bit & n & n=8-bit \\ \hline
			Exact 1 \cite{ref19} & 22n & 176 & 2n+3 & 19 \\ \hline
			Exact 2 \cite{ref34} & 23n & 184 & 2n+3 & 19 \\ \hline
			\multicolumn{5}{|c|}{Scenario 1: five most significant full adders are exact.} \\ \hline
			SIAFA1 \cite{ref3} & 8(n-5)+22(n-3) & 134 & 2n+3 & 19 \\ \hline
			SIAFA2 \cite{ref3} & 10(n-5)+22(n-3) & 140 & 2n+3 & 19 \\ \hline
			SIAFA3 \cite{ref3} & 8(n-5)+22(n-3) & 134 & 2n+3 & 19 \\ \hline
			SIAFA4 \cite{ref3} & 8(n-5)+22(n-3) & 134 & 2n+3 & 19 \\ \hline
			SAFAN \cite{nref1} & 7(n-5)+22(n-3) & 131 & 2n+3 & 19 \\ \hline
			ICIS1 & 6(n-5)+22(n-3) & 128 & 2n+3 & 19 \\ \hline
			ICIS2 & 6(n-5)+22(n-3) & 128 & 2n+3 & 19 \\ \hline
			ICIS3 & 6(n-5)+22(n-3) & 128 & 2n+3 & 19 \\ \hline
			ECIS & 12(n-5)+22(n-3) & 146 & 2n+3 & 19 \\ \hline
			\multicolumn{5}{|c|}{Scenario 2: four most significant full adders are exact.} \\ \hline
			SIAFA1 \cite{ref3} & 8(n-4)+22(n-4) & 120 & 2n+3 & 19 \\ \hline
			SIAFA2 \cite{ref3} & 10(n-4)+22(n-4) & 128 & 2n+3 & 19 \\ \hline
			SIAFA3 \cite{ref3} & 8(n-4)+22(n-4) & 120 & 2n+3 & 19 \\ \hline
			SIAFA4 \cite{ref3} & 8(n-4)+22(n-4) & 120 & 2n+3 & 19 \\ \hline
			SAFAN \cite{nref1} & 7(n-4)+22(n-4) & 116 & 2n+3 & 19 \\ \hline
			ICIS1 & 6(n-4)+22(n-4) & 112 & 2n+3 & 19 \\ \hline
			ICIS2 & 6(n-4)+22(n-4) & 112 & 2n+3 & 19 \\ \hline
			ICIS3 & 6(n-4)+22(n-4) & 112 & 2n+3 & 19 \\ \hline
			ECIS & 12(n-4)+22(n-4) & 136 & 2n+3 & 19 \\ \hline
			\multicolumn{5}{|c|}{Scenario 3: three most significant full adders are exact.} \\ \hline
			SIAFA1 \cite{ref3} & 8(n-3)+22(n-5) & 106 & 2n+3 & 19 \\ \hline
			SIAFA2 \cite{ref3} & 10(n-3)+22(n-5) & 116 & 2n+3 & 19 \\ \hline
			SIAFA3 \cite{ref3} & 8(n-3)+22(n-5) & 106 & 2n+3 & 19 \\ \hline
			SIAFA4 \cite{ref3} & 8(n-3)+22(n-5) & 106 & 2n+3 & 19 \\ \hline
			SAFAN \cite{nref1} & 7(n-3)+22(n-5) & 101 & 2n+3 & 19 \\ \hline
			ICIS1 & 6(n-3)+22(n-5) & 96 & 2n+3 & 19 \\ \hline
			ICIS2 & 6(n-3)+22(n-5) & 96 & 2n+3 & 19 \\ \hline
			ICIS3 & 6(n-3)+22(n-5) & 96 & 2n+3 & 19 \\ \hline
			ECIS & 12(n-3)+22(n-5) & 126 & 2n+3 & 19 \\ \hline	
	\end{tabular}}
	\label{tab13}
\end{table}

In Table \ref{tab13}, the proposed approximate full adders are compared with the approximate and exact full adders proposed in \cite{ref3,ref19,ref34,nref1} by different circuit evaluation criteria. According to the results of Table \ref{tab13}, all exact and approximate n-bit adders need 2n+3 memristors to perform the two-operands addition. The implementation algorithms of approximate full adders ICIS1-3, SIAFA1 \cite{ref3}, SIAFA3 \cite{ref3}, SIAFA4 \cite{ref3}, and SAFAN \cite{nref1} need four memristors (3 input memristors and one work memristor). However, exact full adders applied in MSBs of n-bit approximate RCA structures need two work memristors. So, there is a need for 2n input memristors, a memristor for the $C_{in}$, and two work memristors. Due to this fact, applying the approximate full adders in the LSBs of n-bit approximate RCA structures does not improve the area factor.

The number of computational steps required to calculate the final results of the exact and approximate n-bit adder structures applying the proposed approximate full adders is reported in Table \ref{tab13}. In n-bit approximate RCA structures, $m_{1}$ LSBs are calculated by applying the approximate full adders, and $m_{2}$ MSBs ($m_{2}=n-m_{1}$) are computed applying the exact full adders. The full adder proposed in \cite{ref19} is applied in the MSBs. So the total number of computational steps of an n-bit approximate RCA adder can be calculated by (\ref{eq25}). In (\ref{eq25}), $\alpha$ refers to the number of computational steps of the desired approximate full adder, e.g., $\alpha$ is 6 for ICIS1 and 12 for ECIS. 

\begin{gather}
	Total~No.~of~steps=\alpha (n-m_{2}) + 22(n-m_{1}) \label{eq25}
\end{gather}

The number of computational steps improved by 2\%-17\% compared to the SIAFA1-4 \cite{ref3} and SAFAN \cite{nref1} and 27\%-48\% compared to 8-bit exact adders by applying ICIS1-3 in scenarios 1-3 (See (\ref{eq25})). The number of computational steps increases 4\%-24\% when the ECIS is applied in the structures of scenarios 1-3 instead of ICIS1-3, SIAFA1-4 \cite{ref3}, and SAFAN \cite{nref1}. However, two things should be noted: (1) the accuracy of computations of ECIS is higher than ICIS1-3, SIAFA1-4 \cite{ref3}, and SAFAN \cite{nref1}, and (2) the number of computational steps when ECIS is placed in the 8-bit adders of scenarios 1-3 is reduced by 17\%-32\% compared to the 8-bit exact adders \cite{ref19,ref34}.

Energy consumption is one of the most critical circuit analysis metrics. One of the main goals of applying approximate computing in the design of arithmetic circuits is to reduce energy consumption. The LTSPICE energy calculation tool is applied to calculate the energy consumption of the proposed approximate full adders, and state-of-the-art \cite{ref3,ref19,ref34,nref1}. First, the sum of the energy consumption of the memristors involved in the computations is measured for each input state of the full adder cell \cite{ref3}. The average energy consumption of all the input states of the full adder cell is reported as the energy consumption estimation of the full adder cell in Table \ref{tab14} \cite{ref3}. According to the results of Table \ref{tab14}, ICIS1-3 have almost equal energy consumption. The number of computational steps and memristor switching, considering dynamic energy consumption, directly affect the amount of energy consumed in the stateful memristive arithmetic structures. Based on their implementation algorithms, these three circuits require four memristors to compute their outputs in six computational steps. Their energy consumption is almost equal because of the same number of computational steps and required memristors. The energy consumption of each exact or approximate full adder can be compared with other full adders based on the number of computational steps and the number of memristors required.

\begin{table}[h]
	\centering
	\caption{Approximate and exact full adders energy consumption comparison.}
	\scalebox{0.9}{
		\begin{tabular}{|c|c|c|}
			\hline
			Serial & Energy & Improvement \\
			full adder & Consumption & percentage \\ 
			(exact and approximate) & ($\times 10^{-9} J$) & over \cite{ref34} \\ \hline
			Exact 1 \cite{ref19} & 1.90859 & 5 $\%$ \\ \hline
			Exact 2 \cite{ref34} & 2.00727 & - \\ \hline
			SIAFA1 \cite{ref3} & 0.67221 & 67 $\%$ \\ \hline
			SIAFA2 \cite{ref3} & 0.86032 & 57 $\%$ \\ \hline
			SIAFA3 \cite{ref3} & 0.67221 & 67 $\%$ \\ \hline
			SIAFA4 \cite{ref3} & 0.67086 & 67 $\%$ \\ \hline
			SAFAN \cite{nref1} & 0.64282 & 68 $\%$ \\ \hline
			ICIS1 & 0.50709 & 75 $\%$ \\ \hline
			ICIS2 & 0.50705 & 75 $\%$ \\ \hline
			ICIS3 & 0.50705 & 75 $\%$ \\ \hline
			ECIS & 1.02631 & 49 $\%$ \\ \hline
	\end{tabular}}
	\label{tab14}
\end{table}

The energy consumption improvement of the proposed approximate full adders compared to the exact full adder \cite{ref34} is written in Table \ref{tab14}. ICIS1-3, as the most energy-efficient approximate full adders, improved the energy consumption by 20\%-75\% compared to the state-of-the-art \cite{ref3,ref19,ref34,nref1}. The energy consumption of the ECIS is higher than of the other approximate full adders (ICIS 1-3, SAFAN \cite{nref1}, and SIAFA1-4 \cite{ref3}). However, this approximate full adder increases computations' accuracy compared to the others. The ECIS improved the energy consumption of exact full adders \cite{ref19,ref34} by 46\% and 49\%.

In general, the energy consumption of n-bit approximate adders applying $m_{1}$ exact full adders \cite{ref19} in MSBs and $m_{2}$ approximate ones in LSBs of RCA structure can be estimated by (\ref{eq26}). $\beta$ is equal to the energy consumption of each approximate full adder.

\begin{gather}
	\begin{split}
			Total~energy~dissipation=\beta(n-m_{1})nJ + 1.90859(n-m_{2})nJ \label{eq26}
	\end{split}
\end{gather}

The limitations of the analyses of this study, along with the functionality assessment of proposed approximate full adders in a 1T1R crossbar structure, are briefly discussed at the end of this subsection.  Employing practical models that incorporate various design details, including diverse parasitic features, enhances the quality of the simulation. The authors of this article did not have access to a model based on a specific fabrication technology that accounts for parasitic features in detail, nor to simulation and layout design tools based on these technologies. Accordingly, the model applied in \cite{ref3,ref9,ref13,ref21,nref1,nref3,nref5,nref6} was employed for the simulation to enable a fair comparison between previously verified circuits and the proposed circuits in this article. The authors have also simulated the proposed approximate full adders algorithms in the 1T1R structure in addition to the 1R structure. The main reason for this simulation is to assess the correct functionality of the proposed circuit in this architecture. The use of 1T1R crossbar array architecture can serve as an efficient solution to address design challenges based on 1R crossbar array structures, such as sneak current path \cite{ref9}. Voltages proportional to the time intervals of the pulse time for the proper functionality of the circuits based on the proposed algorithms have been applied to the bit and word lines. The output memristors in this structure have also been specified in accordance with the proposed algorithm.

\subsection{Error analysis simulation and comparison} \label{sec42}
Two parameters should always be evaluated and analyzed in approximate computing. The first one, the circuit analysis metrics improvement, is analyzed in the last subsection. The computation's accuracy, as the second main parameter in approximate computing, is analyzed in this subsection. Estimating the error magnitude of an error-tolerant application's outputs is possible by examining error analysis metrics \cite{ref38}. Error analysis metrics such as MED and NMED can be applied to analyze the proposed circuits’ accuracy. The smaller the error analysis metrics, the higher the computations’ accuracy. All possible input combinations (65536 different input patterns) are applied to the 8-bit approximate adders of scenarios 1-3, designed based on ICIS1-3, ECIS, SIAFA1-4 \cite{ref3}, and SAFAN \cite{nref1} to calculate the error analysis metrics. The results of error analysis metrics are shown in Table \ref{tab15}.

\begin{table}[h]
	\centering
	\caption{Simulation results of error-analysis metrics.}
	\scalebox{0.9}{
		\begin{tabular}{|c|c|c|}
			\hline
			Approximate full adder & MED & NMED \\ \hline
			\multicolumn{3}{|c|}{Scenario 1: five most significant full adders are exact.} \\ \hline
			SIAFA1 \cite{ref3} & 2.062 & 0.004 \\ \hline
			SIAFA2 \cite{ref3} & 2.656 & 0.0052 \\ \hline
			SIAFA3 \cite{ref3} & 2.062 & 0.004 \\ \hline
			SIAFA4 \cite{ref3} & 2.625 & 0.0051 \\ \hline
			SAFAN \cite{nref1} & 2.9375 & 0.0057 \\ \hline
			ICIS1 & 2.156 & 0.0042 \\ \hline
			ICIS2 & 2.25 & 0.0044 \\ \hline
			ICIS3 & 2.25 & 0.0044 \\ \hline
			ECIS & 1.718 & 0.0033 \\ \hline
			\multicolumn{3}{|c|}{Scenario 2: four most significant full adders are exact.} \\ \hline
			SIAFA1 \cite{ref3} & 4.351 & 0.0085 \\ \hline
			SIAFA2 \cite{ref3} & 6.1718 & 0.0121 \\ \hline
			SIAFA3 \cite{ref3} & 4.351 & 0.0085 \\ \hline
			SIAFA4 \cite{ref3} & 5.3125 & 0.0104 \\ \hline
			SAFAN \cite{nref1} & 5.78125 & 0.0113 \\ \hline
			ICIS1 & 4.7265 & 0.0092 \\ \hline
			ICIS2 & 4.4687 & 0.0087 \\ \hline
			ICIS3 & 4.4687 & 0.0087 \\ \hline
			ECIS & 3.6171 & 0.007 \\ \hline
			\multicolumn{3}{|c|}{Scenario 3: three most significant full adders are exact.} \\ \hline
			SIAFA1 \cite{ref3} & 8.8554 & 0.0173 \\ \hline
			SIAFA2 \cite{ref3} & 13.498 & 0.0264 \\ \hline
			SIAFA3 \cite{ref3} & 8.8554 & 0.0173 \\ \hline
			SIAFA4 \cite{ref3} & 10.6562 & 0.0208 \\ \hline
			SAFAN \cite{nref1} & 11.04687 & 0.02166 \\ \hline
			ICIS1 & 9.8886 & 0.0193 \\ \hline
			ICIS2 & 8.9121 & 0.0174 \\ \hline
			ICIS3 & 8.9121 & 0.0174 \\ \hline
			ECIS & 7.3769 & 0.0144 \\ \hline
	\end{tabular}}
	\label{tab15}
\end{table}

The main goal of proposing the ECIS is to prevent the spread of inexact $C_{out}$ from LSBs to the MSBs and increase the accuracy of computations. Error analysis metrics (e.g., MED and NMED) are minimum when the ECIS is applied in the structure of scenarios 1-3, according to the results of Table \ref{tab15}. The accuracy of computations increases when the ECIS full adder is applied in larger arithmetic structures compared to the ICIS1-3, SIAFA1-4 \cite{ref3}, and SAFAN \cite{nref1}, based on the results of Table \ref{tab15}. So, the primary purpose of proposing the ECIS is achieved. The error analysis metrics of ICIS2 and ICIS3 full adders are equal, based on their truth tables, in scenarios 1-3, and the results are reported in Table \ref{tab15}. The ED of these two approximate full adders equals 3, and their computations’ accuracy is lower than ECIS (ED=2). The computations’ accuracy decreased by 9\%, 2\%, and 1\%, respectively, by applying the ICIS2 and ICIS3 instead of SIAFA1 \cite{ref3} and SIAFA3 \cite{ref3} in the structures of scenarios 1-3. It should be mentioned that the number of computational cycles has also decreased by 5\%, 7\%, and 9\% by applying the ICIS2 and ICIS3 instead of SIAFA1 \cite{ref3} and SIAFA3 \cite{ref3} in these three scenarios, respectively. The computations’ accuracy of ICIS1 is lower than the other proposed approximate full adders in this article, SIAFA1 \cite{ref3}, and SIAFA3 \cite{ref3} in scenarios 1-3. The accuracy of computations decreased from 4\%-10\% by applying ICIS1 instead of SIAFA1 \cite{ref3} and SIAFA3 \cite{ref3} in scenarios 1-3, but the computational steps improved by 5\%-9\% in these scenarios. This approximate full adder increases the accuracy of computations in scenarios 1-3 compared to SIAFA2 \cite{ref3}, SIAFA4 \cite{ref3}, and SAFAN \cite{nref1}.

\subsection{Application-level simulation and comparison} \label{sec43}
The reduction of circuit complexity in approximate computing should be accompanied by an acceptable accuracy reduction so that this computational method can be effectively applied in error-resilient applications. Image processing as an error-resilient application is a complex and data-intensive application widely applied in daily human lives  \cite{ref3}. This application is a suitable candidate for in-memory approximate computing by applying memristors \cite{ref3,ref8}. So, the proposed approximate full adders are applied in the computational structures of scenarios 1-3 and are simulated behaviorally in three different image processing applications of image addition, motion detection, and grayscale filter. PSNR, SSIM, and MSSIM are applied to evaluate the output image quality metrics. The proposed full adders have equal single-bit error analysis metrics, but do not have equal output image quality metrics in image processing applications. The reason is that the input distributions in images differ from those in the error analysis simulations \cite{ref3}.

\subsubsection{Image addition application} \label{sec431}
Improving the quality of the images and masking them are the applications of image addition \cite{ref3}. Image addition is a fundamental and widely used application in image processing. In the image addition application, the pixels $P_{1_{ij}}$ and $P_{2_{ij}}$ of the two input images are added together, and the result is stored in the output pixel $P_{O_{ij}}$.

\begin{figure}[hbt!]
	\centering
	\includegraphics[scale=0.22]{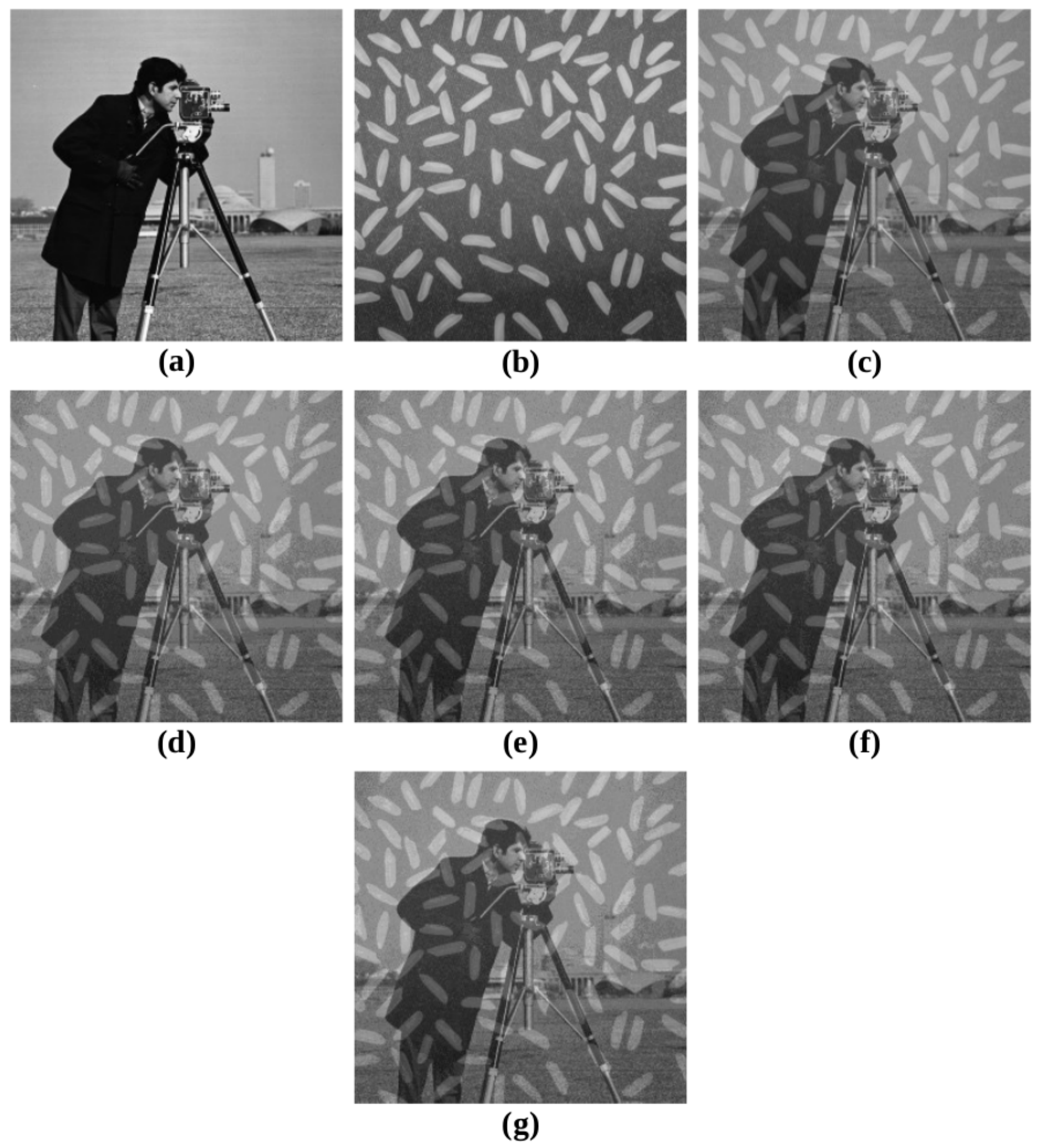}
	\caption{The image addition simulation outputs based on the third scenario's RCA structure: (a) cameraman, (b) rice, (c) exact output, (d) ICIS1, (e) ICIS2, (f) ICIS3, and (g) ECIS.}
	\label{fig7}
\end{figure}

The behavioral simulation results of ECIS and ICIS1-3 that are applied in the image addition application in the computational structures of scenarios 1-3 are tabulated in Table \ref{tab16}. The third scenario’s output images for the proposed approximate full adders (ICIS1-3 and ECIS) are illustrated in Figure \ref{fig7}.

\begin{table}[h]
	\centering
	\caption{Image quality metrics of image addition application.}
	\scalebox{0.75}{
		\begin{tabular}{|c|c|c|c|}
			\hline
			Approximate & PSNR & SSIM & MSSIM \\
			full adder & (dB) & & \\ \hline
			\multicolumn{4}{|c|}{Scenario 1: five most significant full adders are exact.} \\ \hline
			SIAFA1 \cite{ref3} & 44.5148 & 0.9899 & 0.99 \\ \hline
			SIAFA2 \cite{ref3} & 41.9674 & 0.9858 & 0.9861 \\ \hline
			SIAFA3 \cite{ref3} & 44.5222 & 0.9898 & 0.99 \\ \hline
			SIAFA4 \cite{ref3} & 43.7483 & 0.9878 & 0.988 \\ \hline
			SAFAN \cite{nref1} & 41.8917 & 0.994 & 0.994 \\ \hline
			ICIS1 & 44.1644 & 0.9909 & 0.991 \\ \hline
			ICIS2 & 43.9423 & 0.9888 & 0.9889 \\ \hline
			ICIS3 & 43.9769 & 0.9886 & 0.9887 \\ \hline
			ECIS & 45.1444 & 0.9918 & 0.9919 \\ \hline
			\multicolumn{4}{|c|}{Scenario 2: four most significant full adders are exact.} \\ \hline
			SIAFA1 \cite{ref3} & 38.67 & 0.9644 & 0.9649 \\ \hline
			SIAFA2 \cite{ref3} & 35.4576 & 0.9425 & 0.9436 \\ \hline
			SIAFA3 \cite{ref3} & 38.8399 & 0.9638 & 0.9644 \\ \hline
			SIAFA4 \cite{ref3} & 37.8083 & 0.959 & 0.9597 \\ \hline
			SAFAN \cite{nref1} & 36.6395 & 0.9793 & 0.9796 \\ \hline
			ICIS1 & 38.2287 & 0.9654 & 0.966 \\ \hline
			ICIS2 & 38.545 & 0.9632 & 0.9636 \\ \hline
			ICIS3 & 38.4096 & 0.961 & 0.9615 \\ \hline
			ECIS & 39.4711 & 0.9702 & 0.9706 \\ \hline
			\multicolumn{4}{|c|}{Scenario 3: three most significant full adders are exact.} \\ \hline
			SIAFA1 \cite{ref3} & 32.9823 & 0.8974 & 0.8996 \\ \hline
			SIAFA2 \cite{ref3} & 28.2504 & 0.8156 & 0.8166 \\ \hline
			SIAFA3 \cite{ref3} & 32.6497 & 0.8905 & 0.8915 \\ \hline
			SIAFA4 \cite{ref3} & 32.0442 & 0.8931 & 0.8956 \\ \hline
			SAFAN \cite{nref1} & 30.5866 & 0.9297 & 0.9298 \\ \hline
			ICIS1 & 32.0474 & 0.9006 & 0.9027 \\ \hline
			ICIS2 & 32.9714 & 0.896 & 0.8978 \\ \hline
			ICIS3 & 33.0242 & 0.8927 & 0.8956 \\ \hline
			ECIS & 33.7765 & 0.9128 & 0.9143 \\ \hline
	\end{tabular}}
	\label{tab16}
\end{table}

According to Table \ref{tab16}, the highest quality of the output images is obtained when ECIS is applied in the computational structures of scenarios 1-3 introduced in subsection \ref{sec41}. The lowest quality of the output images (PSNR) corresponded to SIAFA2 \cite{ref3} and SAFAN \cite{nref1}. Also, the quality of output images is acceptable when ICIS1-3 are applied in the computational structures of scenarios 1-3. The output image quality is higher than 45 dB when the ECIS is applied in scenario 1. The quality of the output images varied between 43.94 dB and 44.5 dB when ICIS1-3, SIAFA1 \cite{ref3}, and SIAFA3 \cite{ref3} are applied in the computational structure of scenario 1. The quality of the output images is higher than 39 dB only when ECIS is applied in the second scenario’s computing structure. The quality of the output images is between 38-39 dB when ICIS1-3, SIAFA1 \cite{ref3}, and SIAFA3 \cite{ref3} are applied in the second scenario’s RCA structure. The outputs of the image addition application have the least image quality when SAFAN \cite{nref1} and SIAFA2 \cite{ref3} are applied in the second scenario’s arithmetic architecture (PSNR is less than 38 dB). The quality of output images is higher than of the other circuits by applying ECIS and ICIS3 in the third scenario (PSNR is higher than 33 dB). The outputs are acceptable when ICIS1, ICIS2, SIAFA1 \cite{ref3}, SIAFA3 \cite{ref3}, SIAFA4 \cite{ref3}, and SAFAN \cite{nref1} are applied in the LSBs of the third scenario’s approximate adder. Suppose the number of approximate full adders (ECIS, ICIS1-3, SIAFA1-4 \cite{ref3}, and SAFAN \cite{nref1}) increases in the LSBs of an 8-bit RCA to six bits. In that case, the image quality of the outputs generated by these RCAs is less than 30 dB, and the output images are unacceptable.

\subsubsection{Motion Detection Application} \label{sec432}
Subtraction of two images is applied to detect motion in images \cite{ref3}. The corresponding pixels of two images are subtracted from each other, and the result is stored as an output image. In this article, the 2’s complement addition method is applied in the approximate adders of scenarios 1-3 to perform subtraction.

\begin{figure}[hbt!]
	\centering
	\includegraphics[scale=0.22]{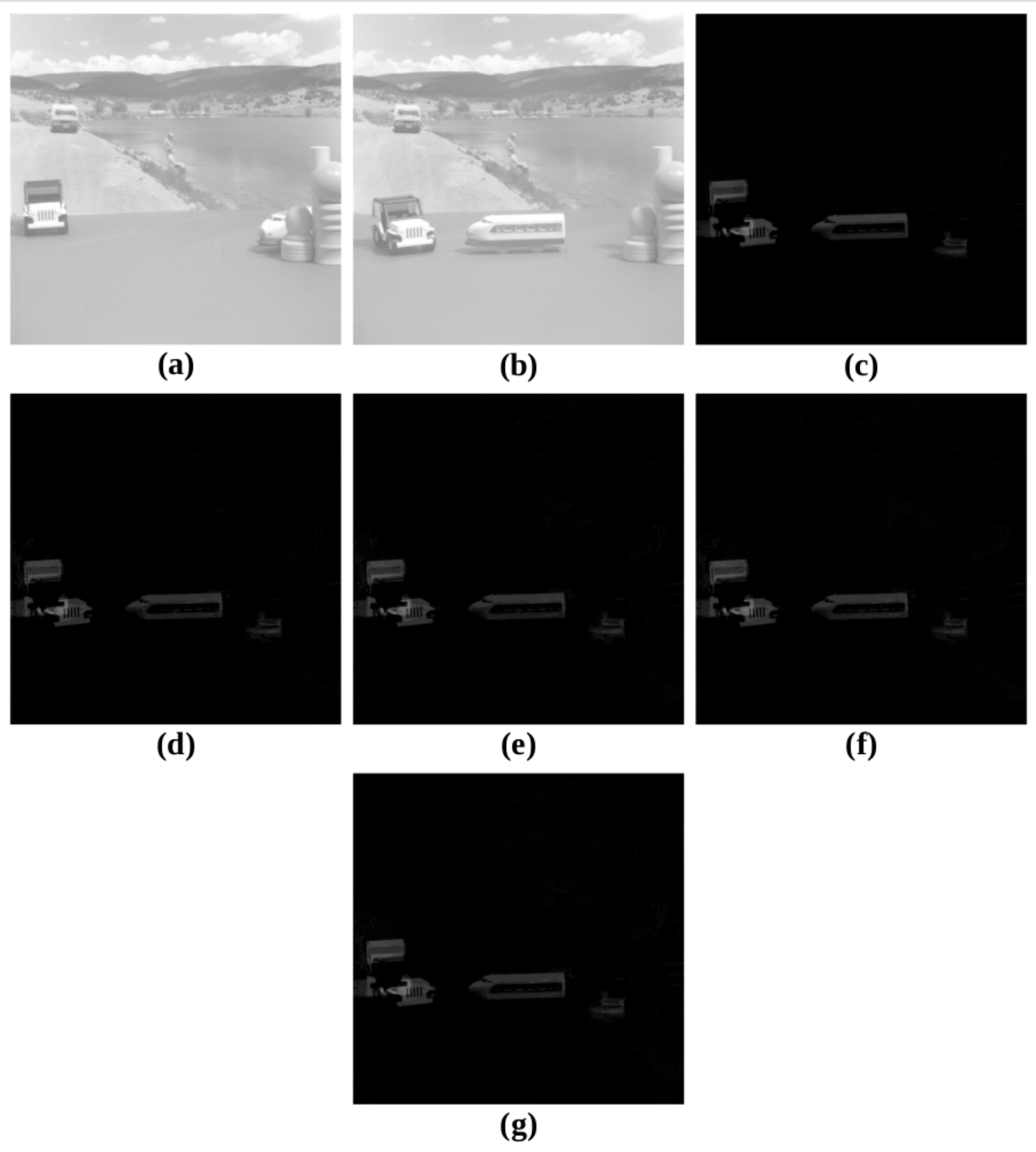}
	\caption{The motion detection simulation outputs based on the second scenario's RCA structure: (a) first image \cite{ref39}, (b) second image \cite{ref39}, (c) exact output, (d) ICIS1, (e) ICIS2, (f) ICIS3, and (g) ECIS.}
	\label{fig8}
\end{figure}

Two consecutive frames are placed in the subfigures \ref{fig8}(a) and \ref{fig8}(b). The result of the subtraction of these two consecutive frames, applying an exact 8-bit adder is shown in the subfigure \ref{fig8}(c). Subfigures \ref{fig8}(d)-\ref{fig8}(g) are the output images obtained from the subtraction of these two consecutive frames applying the proposed approximate full adders in the RCA of scenario 2. The image quality metrics of the output images in different scenarios are written in Table \ref{tab17}.

\begin{table}[h]
	\centering
	\caption{Image quality metrics of motion detection application.}
	\scalebox{0.75}{
		\begin{tabular}{|c|c|c|c|}
			\hline
			Approximate & PSNR & SSIM & MSSIM \\
			full adder & (dB) & & \\ \hline
			\multicolumn{4}{|c|}{Scenario 1: five most significant full adders are exact.} \\ \hline
			SIAFA1 \cite{ref3} & 41.5919 & 0.7711 & 0.7901 \\ \hline
			SIAFA2 \cite{ref3} & 43.5407 & 0.9678 & 0.9874 \\ \hline
			SIAFA3 \cite{ref3} & 41.8705 & 0.7905 & 0.8103 \\ \hline
			SIAFA4 \cite{ref3} & 45.5121 & 0.9534 & 0.9693 \\ \hline
			SAFAN \cite{nref1} & 49.7504 & 0.9863 & 0.9905 \\ \hline
			ICIS1 & 45.564 & 0.9703 & 0.9886 \\ \hline
			ICIS2 & 45.8315 & 0.9685 & 0.9863 \\ \hline
			ICIS3 & 45.8356 & 0.9682 & 0.9861 \\ \hline
			ECIS & 46.1802 & 0.9661 & 0.9811 \\ \hline
			\multicolumn{4}{|c|}{Scenario 2: four most significant full adders are exact.} \\ \hline
			SIAFA1 \cite{ref3} & 37.4131 & 0.6405 & 0.6705\\ \hline
			SIAFA2 \cite{ref3} & 37.5605 & 0.9338 & 0.9652 \\ \hline
			SIAFA3 \cite{ref3} & 36.8613 & 0.5993 & 0.6302 \\ \hline
			SIAFA4 \cite{ref3} & 40.3861 & 0.9131 & 0.9423 \\ \hline
			SAFAN \cite{nref1} & 44.0205 & 0.9798 & 0.9857 \\ \hline
			ICIS1 & 40.3774 & 0.9409 & 0.9713 \\ \hline
			ICIS2 & 40.5783 & 0.9351 & 0.968 \\ \hline
			ICIS3 & 40.715 & 0.9349 & 0.9685 \\ \hline
			ECIS & 40.7888 & 0.9303 & 0.9588 \\ \hline
			\multicolumn{4}{|c|}{Scenario 3: three most significant full adders are exact.} \\ \hline
			SIAFA1 \cite{ref3} & 32.6121 & 0.508 & 0.5404 \\ \hline
			SIAFA2 \cite{ref3} & 31.6441 & 0.8991 & 0.9265 \\ \hline
			SIAFA3 \cite{ref3} & 32.4096 & 0.4747 & 0.5094 \\ \hline
			SIAFA4 \cite{ref3} & 35.0436 & 0.8664 & 0.902 \\ \hline
			SAFAN \cite{nref1} & 37.5336 & 0.9667 & 0.9727 \\ \hline
			ICIS1 & 34.928 & 0.9104 & 0.9421 \\ \hline
			ICIS2 & 35.2479 & 0.8956 & 0.9363 \\ \hline
			ICIS3 & 35.3686 & 0.8959 & 0.937 \\ \hline
			ECIS & 35.3095 & 0.8873 & 0.9226 \\ \hline
	\end{tabular}}
	\label{tab17}
\end{table}

According to the results of Table \ref{tab17}, ECIS and ICIS1-3 are among the first six approximate full adders in the application of motion detection in all three scenarios. When the proposed circuits are applied in the computational structures of scenarios 1-3, the quality of the output images is higher than 45 dB, 40 dB, and 34.9 dB, respectively. The image quality metrics of the outputs of motion detection application are not acceptable if the number of approximate full adders in the LSBs of an 8-bit approximate adder structure is changed from 5 to 6 (only the two MSBs are computed applying the exact full adder). The PSNR of output images is higher than 29.5 dB, but the motions could not be detected by applying all four proposed approximate full adders in this structure. The same results are obtained by applying the state-of-the-art \cite{ref3, nref1} in this RCA, and it is impossible to detect motion by applying these approximate full adders.

\subsubsection{Grayscale filter} \label{sec433}
Red, Green, and Blue (RGB) images are made of an $m \times n \times 3$ vector where $m \times n$ defines the image's dimensions. Each RGB image consists of three layers R, G, and B. Combining these three layers in different ways creates an RGB image. In this paper, like \cite{ref3, nref1}, each pixel's R, G, and B values are added together to convert RGB images to grayscale by the RCAs of scenarios 1-3 designed based on ECIS, ICIS1-3, SIAFA1-4 \cite{ref3}, and SAFAN \cite{nref1}. Then the sum of these values is divided by three and stored as the output grayscale pixel. It should be noted that the division is computed accurately \cite{ref3}.

\begin{figure}[hbt!]
	\centering
	\includegraphics[scale=0.22]{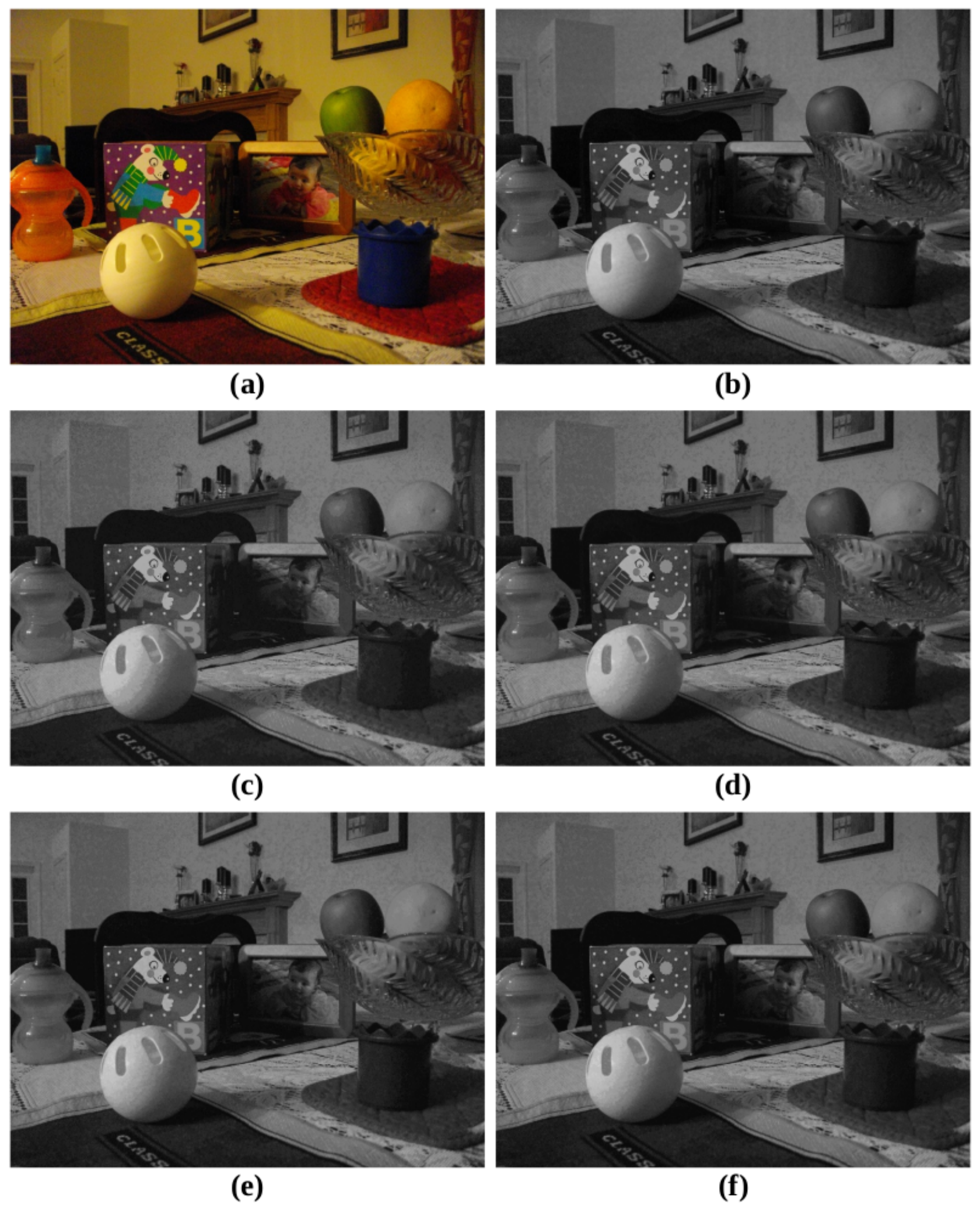}
	\caption{The grayscale filter simulation outputs based on the third scenario's RCA structure: (a) RGB image, (b) exact output, (c) ICIS1, (d) ICIS2, (e) ICIS3, and (f) ECIS.}
	\label{fig9}
\end{figure}

The simulation results of this image processing application applying the proposed approximate full adders in the computational structures of scenarios 1-3 are written in Table \ref{tab18}. Figure \ref{fig9} shows the grayscale output images computed by the approximate RCAs of scenario three made by the proposed circuits.

\begin{table}[hbt!]
	\centering
	\caption{Image quality metrics of grayscale filter.}
	\scalebox{0.75}{
		\begin{tabular}{|c|c|c|c|}
			\hline
			Approximate & PSNR & SSIM & MSSIM \\
			full adder & (dB) & & \\ \hline
			\multicolumn{4}{|c|}{Scenario 1: five most significant full adders are exact.} \\ \hline
			SIAFA1 \cite{ref3} & 47.1982 & 0.9911 & 0.999 \\ \hline
			SIAFA2 \cite{ref3} & 43.1339 & 0.9833 & 0.9977 \\ \hline
			SIAFA3 \cite{ref3} & 47.2496 & 0.9914 & 0.999 \\ \hline
			SIAFA4 \cite{ref3} & 43.0565 & 0.9841 & 0.997 \\ \hline
			SAFAN \cite{nref1} & 42.2449 & 0.9905 & 0.9974 \\ \hline
			ICIS1 & 44.7574 & 0.9878 & 0.9984 \\ \hline
			ICIS2 & 45.4874 & 0.9911 & 0.9987 \\ \hline
			ICIS3 & 44.9761 & 0.9903 & 0.9986 \\ \hline
			ECIS & 47.5379 & 0.9925 & 0.9991 \\ \hline
			\multicolumn{4}{|c|}{Scenario 2: four most significant full adders are exact.} \\ \hline
			SIAFA1 \cite{ref3} & 41.4201 & 0.9693 & 0.9957 \\ \hline
			SIAFA2 \cite{ref3} & 35.9998 & 0.9263 & 0.9874 \\ \hline
			SIAFA3 \cite{ref3} & 41.2315 & 0.9684 & 0.9956 \\ \hline
			SIAFA4 \cite{ref3} & 36.9634 & 0.9451 & 0.989 \\ \hline
			SAFAN \cite{nref1} & 36.0101 & 0.9625 & 0.9886 \\ \hline
			ICIS1 & 37.9719 & 0.9474 & 0.9911 \\ \hline
			ICIS2 & 40.1764 & 0.9693 & 0.9953 \\ \hline
			ICIS3 & 39.9446 & 0.9679 & 0.9952 \\ \hline
			ECIS & 41.9064 & 0.973 & 0.9966 \\ \hline
			\multicolumn{4}{|c|}{Scenario 3: three most significant full adders are exact.} \\ \hline
			SIAFA1 \cite{ref3} & 35.5671 & 0.9019 & 0.9778 \\ \hline
			SIAFA2 \cite{ref3} & 28.4883 & 0.7576 & 0.9317 \\ \hline
			SIAFA3 \cite{ref3} & 35.3588 & 0.8916 & 0.9794 \\ \hline
			SIAFA4 \cite{ref3} & 31.5146 & 0.8525 & 0.9589 \\ \hline
			SAFAN \cite{nref1} & 28.9472 & 0.8633 & 0.9482 \\ \hline
			ICIS1 & 30.3889 & 0.8163 & 0.9441 \\ \hline
			ICIS2 & 33.9817 & 0.9011 & 0.973 \\ \hline
			ICIS3 & 34.4268 & 0.8992 & 0.9782 \\ \hline
			ECIS & 35.8643 & 0.9094 & 0.9814 \\ \hline
	\end{tabular}}
	\label{tab18}
\end{table}

The output images have the best image quality metrics when ECIS is applied in scenarios 1-3. The output images of other proposed approximate full adders (ICIS1-3) are also acceptable in scenarios 1-3. In the first, second, and third scenarios, the PSNR of the output images generated by the proposed approximate full adders is higher than 44.7 dB, 37.9 dB, and 30 dB, respectively.

The simulation result of this application is also of good quality, even though a small shadow is formed on the output image if the ECIS is applied in the six LSBs of an 8-bit approximate RCA (two MSBs are exact). The PSNR of the output image in this scenario is 29.8814 dB. The output’s image quality metrics are unacceptable when the ICIS1 is applied in the six LSBs of an 8-bit approximate RCA (PSNR is 23.4527 dB). The output images and the grayscale filter's performance are unacceptable (PSNR is about 27 dB) when ICIS2 and ICIS3 are also applied in the mentioned structure.

\subsection{Analysis of the trade-off between Circuit and Error evaluation metrics} \label{sec44}
The proposed circuits were analyzed only by circuit analysis metrics in subsection \ref{sec41}. Different error analysis criteria and image quality metrics were applied to analyze the proposed circuits in subsections \ref{sec42} and \ref{sec43}. The reduction of accuracy in computations leads to the decrease in circuit complexity in approximate computing. We compared the proposed circuits by simultaneously considering the circuit and error evaluation criteria in this subsection. The $FOM_{1}$ introduced in this article considers the $\underline{E}$nergy consumption and computational $\underline{D}$elay $\underline{P}$roduct (EDP) as circuit analysis metrics and the accuracy-related metrics of 1-NMED and PSNR according to the third scenario's results.

This $FOM_{1}$ is defined as
\begin{gather}
	FOM_{1}=\frac{EDP}{(1-NMED) \times PSNR} \label{eq27}
\end{gather}

The results of $FOM_{1}$ for the proposed circuits, SIAFA1-4 \cite{ref3}, and SAFAN \cite{nref1} are presented in Table \ref{tab19}. The average PSNR value of three 8-bit image processing applications assessed in the last subsection is considered to calculate the $FOM_{1}$. ICIS3, ICIS2, and ICIS1 made the best trade-off between the circuit complexity and the accuracy of the computations, respectively, according to Table \ref{tab19}. The $FOM_{1}$ computed for the ECIS is higher than ICIS1-3, SAFAN \cite{nref1}, SIAFA1 \cite{ref3}, SIAFA3 \cite{ref3}, and SIAFA4 \cite{ref3}. However, it should be emphasized that the $ER_{C_{out}}$ of ECIS is equal to 0, and errors do not propagate from LSBs to MSBs. This full adder's computations’ accuracy is higher than the other approximate full adders.

The applicability of ECIS is recommended over the other proposed full adders in applications where the accuracy of computations is of great importance and a limited reduction of hardware complexity is acceptable.

\begin{table}[h]
	\centering
	\caption{The $FOM_{1}$ results of the SIAFA1-4 \cite{ref3}, SAFAN \cite{nref1}, and the proposed approximate full adders in the third scenario's RCA structure.}
	\scalebox{0.9}{
		\begin{tabular}{|c|c|}
			\hline
			Approximate full adder & $FOM_{1}$ \\ \hline
			SIAFA1 \cite{ref3} & 29.06734 \\ \hline
			SIAFA2 \cite{ref3} & 40.55279 \\ \hline
			SIAFA3 \cite{ref3} & 29.2824 \\ \hline
			SIAFA4 \cite{ref3} & 29.9061 \\ \hline
			SAFAN \cite{nref1} & 28.61973 \\ \hline
			ICIS1 & 24.91743  \\ \hline
			ICIS2 & 23.69175  \\ \hline 
			ICIS3 & 23.54914  \\ \hline 
			ECIS & 39.67634  \\ \hline 
	\end{tabular}}
	\label{tab19}
\end{table}

Energy saving is of great importance in the design of approximate adders. Examining the trade-off between accuracy, energy saving criteria, and improving computational delay (delay saving) in Error-tolerant applications can provide a suitable quantification between the basic objectives of approximate computing. Accordingly, $FOM_{2}$ has been investigated to examine this trade-off based on what has been proposed in \cite{nref7}. In this criterion, power saving percentage, delay saving percentage, and $PSNR^{2}$ have been applied to simultaneously analyze the improvement of circuit evaluation criteria and approximate output quality (see (\ref{eqfom2})). The results of $FOM_{2}$ computation for the third scenarios’ approximate adder are written in Table \ref{tab20}.

\begin{gather}
	FOM_{2}=Energy~saving \times Delay~ saving \times PSNR^{2} \label{eqfom2}
\end{gather}

\begin{table}[h]
	\centering
	\caption{The $FOM_{2}$ results of the SIAFA1-4 \cite{ref3}, SAFAN \cite{nref1}, and the proposed approximate full adders in the third scenario's RCA structure.}
	\scalebox{0.9}{
		\begin{tabular}{|c|c|}
			\hline
			Approximate full adder & $FOM_{2}$ \\ \hline
			SIAFA1 \cite{ref3} & 183.102 \\ \hline
			SIAFA2 \cite{ref3} & 101.546 \\ \hline
			SIAFA3 \cite{ref3} & 180.42 \\ \hline
			SIAFA4 \cite{ref3} & 174.169 \\ \hline
			SAFAN \cite{nref1} & 184.944 \\ \hline
			ICIS1 & 219.784  \\ \hline
			ICIS2 & 242.164  \\ \hline 
			ICIS3 & 245.104  \\ \hline 
			ECIS & 100.483  \\ \hline 
	\end{tabular}}
	\label{tab20}
\end{table}

According to the results of Table \ref{tab20}, it can be concluded that ICIS3, ICIS2, and ICIS1 have made the best trade-off between circuit analysis metrics and accuracy in image processing applications.

\section{Conclusion} \label{sec5}
We tried to take a step forward to improve the power wall problem by applying approximate computing. The ICIS1-3 are proposed to improve energy consumption by applying approximate computing and redefining approximate logic from the exact logic method. The ECIS is also proposed to increase the accuracy of computations while improving the circuit analysis metrics compared to the exact full adders. The implementation algorithms of the proposed full adders are designed by applying the IMPLY method as a stateful logic to enable the implementation of these circuits for IMP in the structure of crossbar arrays. ICIS1-3 improve the energy consumption by 21\%-41\% over approximate state-of-the-art and 73\%-75\% over the exact full adders. ICIS1-3 improve the computational steps by 14\%-74\% compared to state-of-the-art, applying the same number of memristors as previous designs. The ECIS also reduces energy consumption by a maximum of 49\% and the number of computational steps by 48\% compared to exact full adders. The proposed full adders are compared with state-of-the-art by error analysis metrics and image quality criteria in three scenarios and three image processing applications. The ECIS has the highest computation accuracy in all three scenarios. ICIS1-3 have acceptable error evaluation criteria in all three scenarios and image processing applications. Circuit evaluation metrics and accuracy in computations are evaluated side by side by presenting two $FOMs$. ICIS3, ICIS2, and ICIS1 have the highest ranks among other approximate full adders concerning the $FOM_{1}$ and $FOM_{2}$, respectively.

\bibliographystyle{ieeetr} 
\bibliography{Manuscript}

\appendix
\renewcommand{\thetable}{A\arabic{table}}
\setcounter{table}{0}

\section{Supplementary application-level simulation} \label{appendixa}
Appendix \ref{appendixa} reports the results of the supplementary application-level simulation (image addition and grayscale filter). The purpose of this simulation was to gain more confidence in the applicability of the proposed circuits (ICIS1-3 and ECIS) in the above-mentioned applications. Accordingly, the proposed approximate full adders were applied in the third scenario structure, which has the highest degree of approximation among the others. Ten grayscale images in the image addition application and five RGB images in the grayscale filter application were applied to the approximate adder of the third scenario, consisting of the proposed approximate full adders. The image quality evaluation criteria were recorded separately for each circuit, and the average values obtained from the simulations were calculated for each criterion. The results are reported in Tables \ref{taba1} (image addition application) and \ref{taba2} (grayscale filter).

\begin{table}[h]
	\centering
	\caption{Image quality metrics of supplementary application-level simulation: image addition application.}
	\scalebox{0.75}{
		\begin{tabular}{|c|c|c|c|}
			\hline
			Approximate & PSNR & SSIM & MSSIM \\
			full adder & (dB) & & \\ \hline
			\multicolumn{4}{|c|}{Scenario 3: three most significant full adders are exact.} \\ \hline
			ICIS1 & 31.8257 & 0.8941 & 0.8955 \\ \hline
			ICIS2 & 32.8602 & 0.8881 & 0.8898 \\ \hline
			ICIS3 & 32.986 & 0.8898 & 0.8908 \\ \hline
			ECIS & 33.5536 & 0.9064 & 0.9051 \\ \hline
	\end{tabular}}
	\label{taba1}
\end{table}

\begin{table}[h]
	\centering
	\caption{Image quality metrics of supplementary application-level simulation: grayscale filter.}
	\scalebox{0.75}{
		\begin{tabular}{|c|c|c|c|}
			\hline
			Approximate & PSNR & SSIM & MSSIM \\
			full adder & (dB) & & \\ \hline
			\multicolumn{4}{|c|}{Scenario 3: three most significant full adders are exact.} \\ \hline
			ICIS1 & 30.8058 & 0.8262 & 0.886 \\ \hline
			ICIS2 & 34.14 & 0.8957 & 0.9337 \\ \hline
			ICIS3 & 34.0658 & 0.8886 & 0.9328 \\ \hline
			ECIS & 35.7012 & 0.9031 & 0.941 \\ \hline
	\end{tabular}}
	\label{taba2}
\end{table}

Based on the results reported in Tables \ref{taba1} and \ref{taba2}, it can be concluded that all four proposed circuits exhibit appropriate computational accuracy when applied to the third scenario structure, and the outputs produced in image addition and grayscale filter applications are of suitable quality. Considering Table \ref{taba1} and comparing it with Table \ref{tab16}, it can be concluded that the outputs obtained from applying approximate full adders ICIS1-3 in the third scenario structure follow the same trends in terms of PSNR in the image addition application. In the grayscale filter application, similar behaviors are observed in the approximate full adders ICIS2 and ICIS3, and their image quality is comparable, as shown in Table \ref{tab18}. This trend also applies to Table \ref{taba2}. ICIS1 and ECIS have the lowest and highest image quality, respectively, according to Tables \ref{tab18} and \ref{taba2}, and the position of both remains the same according to the simulations performed.

\section*{Statements \& Declarations}
\subsection*{Funding}
No funding was received for this research.

\subsection*{Competing interests}
The authors have no relevant financial or non-financial interests to disclose.

\subsection*{Author Contributions}
\noindent \textbf{Seyed Erfan Fatemieh}: Conceptualization, Methodology, Software, Validation, Formal Analysis, Investigation, Data Curation, Writing - Original Draft, Writing - Review \& Editing, Visualization.

\noindent \textbf{Mohammad Reza Reshadinezhad}: Software, Validation, Formal Analysis, Investigation, Resources, Writing - Review \& Editing, Supervision, Project Administration. 

\subsection*{Data Availability Statement}
Data sharing is not applicable to this article as no new data were created or analyzed in this study.

\end{document}